\documentclass[preprint,12pt,authoryear]{elsarticle}

\usepackage{amssymb}
\usepackage{amsmath}

\usepackage{booktabs}  
\usepackage{array}     
\usepackage{makecell}  
\usepackage{color,soul}
\usepackage{enumitem}
\usepackage{comment}
\usepackage{linguex} 
\usepackage{bibentry} 
\usepackage{natbib}
\usepackage{comment}
\usepackage{etoolbox}
\usepackage{multibib}
\usepackage{xcolor}
\usepackage{tipa} 
\usepackage{rotating}
\usepackage{tabularx}
\usepackage{mdframed}
\usepackage{subcaption} 

\definecolor{Acolor}{rgb}{.32,.59,.84}  
\usepackage{bbding}
\usepackage{ragged2e}
\usepackage{adjustbox}
\newcolumntype{P}[1]{>{\RaggedRight\hspace{0pt}}p{#1}}
\usepackage{hanging}
\usepackage{float}
\usepackage{subcaption}
\usepackage[T1]{fontenc}
\usepackage[utf8]{inputenc}
\usepackage{multirow}

\usepackage{xurl}

\journal{Language Sciences}

\begin{document}

\begin{frontmatter}



\title{``It Would Make a Geordie Feel at Home'': Regional Accent Perceptions and Speech Accommodation in Voice AI}
\author[label1]{Dana Serditova\corref{cor1}}
\affiliation[label1]{organization={University of Regensburg},
            addressline={Faculty of Languages, Literature, and Cultures, Department of English and American Studies},
            city={Regensburg},
            postcode={93040},
            state={Bavaria},
            country={Germany}}
\cortext[cor1]{Corresponding author at: University of Regensburg, Faculty of Languages, Literature, and Cultures, Department of English and American Studies, Regensburg, Germany.\\
E-mail address: dana.serditova@ur.de (Dana Serditova).}

\author[label2,label3]{Kevin Tang}
\affiliation[label2]{organization={Heinrich Heine University D\"{u}sseldorf},
            addressline={Faculty of Arts and Humanities, Institute of English and American Studies, Department English Language and Linguistics},
            city={D\"{u}sseldorf},
            postcode={40225},
            state={North Rhine-Westphalia},
            country={Germany}}
\affiliation[label3]{organization={University of Florida},
            addressline={Department of Linguistics},
            city={Gainesville},
            postcode={32611-5454},
            state={Florida},
            country={U.S.A.}}
\author[label4]{Jochen Steffens}
\affiliation[label4]{organization={University of Applied Sciences D\"{u}sseldorf},
            addressline={Faculty of Media, Institute of Sound and Vibration Engineering (ISAVE)},
            city={D\"{u}sseldorf},
            postcode={40476},
            state={North Rhine-Westphalia},
            country={Germany}}


\begin{abstract}
Automatic speech recognition (ASR) systems often perform less accurately for speakers of regional dialects, but less is known about how users respond to these limitations and how regional voices produced by voice assistants (VAs) are evaluated through a sociolinguistic lens. This study investigates these issues among speakers of Newcastle English (Geordies), examining both linguistic accommodation to voice technologies and perceptions of regional versus mainstream-accented VA voices. Participants interacted with a simulated VA that spoke either Newcastle English or Standard Southern British English (SSBE). They subsequently completed semi-structured interviews about their experiences with the experimental VA and voice technologies, language use, accent attitudes, and place orientation. The findings reveal a two-way process of sociolinguistic bias. The Geordie VA was associated with familiarity, warmth, personality, and local belonging, but also with concerns about professionalism and authority, whereas the SSBE voice was perceived as more neutral, clear, and credible. Participants also reported adapting their speech when voice technologies failed to recognise them, most commonly by shifting to a more mainstream accent, enunciating more clearly, and speaking more slowly. These strategies are similar to accommodation in human–human interaction and reflect users’ perceptions of what technologies can recognise, also known as perceived intelligibility pressures. Overall, the findings demonstrate that voice technologies can reproduce existing accent biases both by encouraging users to suppress regional features and by activating established social stereotypes through synthetic voices. The study positions sociolinguistics as a bridge between understanding human–AI interaction and designing inclusive technologies around users’ own linguistic priorities.
\end{abstract}


\begin{keyword}
\def\sep{\unskip;\space}
bias in speech technology
\sep sociolinguistics and AI
\sep speech accommodation
\sep communication accommodation theory
\sep voice assistant
\sep Newcastle English
\end{keyword}




\end{frontmatter}



\section{Introduction}
\label{sec1}

Voice Assistants (VAs) and Automatic Speech Recognition (ASR) have become embedded in many aspects of contemporary life. They support applications in educational contexts \citep{cumbal2024you, butler2019exploration}, healthcare \citep{latif2020speech, adedeji2024sound}, professional communication \citep{zhao2024wildchat, bassignana2025ai}, and everyday communication \citep{vacher2010complete, hamalainen2015multilingual, liang2026wasn}. At the same time, a growing body of research has shown that the benefits of these technologies are not distributed equally across speaker populations. Recognition accuracy is often substantially lower for speakers whose dialects differ from the linguistic mainstream that dominates ASR training data \citep{koenecke2020racial, wassink2022uneven, markl2022language}. Such findings have fuelled discussions surrounding algorithmic fairness \citep{liu2022towards}, equitable access to speech technologies \citep{green2021automatic}, and the sociolinguistic ideologies embedded within computational models of speech \citep{markl2021context, grieve2025sociolinguistic}. Recent scholarship has argued that these disparities cannot be understood solely as engineering challenges. They are the result of the linguistic variation found in everyday speech and and the inability of ASR to handle it \citep[e.g.,][]{BeraAgarwal2025, dipto-etal-2025-asr, ngueajio2022hey, MartinCorpusData2021}.

While previous research has primarily examined recognition accuracy and demographic disparities, less attention has been paid to speakers' own responses to these limitations. If voice technologies consistently fail to recognise particular varieties of speech, users may adapt by modifying their pronunciation, lexical choices, or accent in order to be understood \citep[cf.][]{mengesha2021don}. Such adaptations also reflect speakers' awareness of ``standard'' speech and the social meanings associated with different ways of speaking. Thus, interactions with voice AI provide a novel setting in which established sociolinguistic processes, including accommodation, style-shifting, and language ideologies, can be studied.

This paper investigates these questions through the experiences of speakers of Newcastle English, a salient regional variety of North East England that remains largely underrepresented in contemporary speech technologies \citep{markl2022language,serditova25_interspeech,serditova2026sociolinguistic}. The paper draws on semi-structured interviews conducted as part of the ADAPT-VA project\footnote{Url redacted for double-blind peer review.}, which explores sociolinguistic adaptation in human--computer interaction. We examine how Newcastle English speakers, also known as Geordies, interact with a VA that either speaks with their accent (Newcastle English) or Standard Southern British English (SSBE). By means of a semi-structured post-hoc interview, we further investigate participants' evaluations of regional and standard-accented AI voices and prior experiences with speech technology. We also explore such topics as their attitudes towards Newcastle, connectedness with their communities, and perceptions of their own accent.

This paper integrates research on ASR bias, communication accommodation, and language attitudes and examines two complementary aspects of sociolinguistic interaction with VAs. First, it explores how users evaluate a regional, dialectal VA voice as opposed to a mainstream one, particularly how it is described in terms of familiarity, authenticity, warmth, professionalism, and authority. We then look at how these evaluations compare with and reflect general attitudes towards regional vs. mainstream accents. Second, it investigates how speakers report accommodating to existing speech technologies by modifying their pronunciation or lexical choices in order to achieve successful recognition. We argue that VAs reproduce sociolinguistic biases from both directions: users adapt their speech to technologies perceived to privilege standard dialects, while regional synthetic voices continue to be interpreted through established social evaluations and stereotypes about accents. We ask the following research questions:

\begin{enumerate}
    \item[\textbf{RQ1.}] How do Geordie speakers evaluate regional and mainstream-accented VA voices, and how do these evaluations reflect general sociolinguistic attitudes towards accent?

    \item[\textbf{RQ2.}] How do Geordie speakers report accommodating their speech to voice technologies, and how does this accommodation relate to the ways they modify their speech in everyday interactions?
\end{enumerate}

\section{Previous Research}
\label{sec2}

This section reviews previous research on sociolinguistic biases in ASR, linguistic accommodation in human--computer interaction, and social perceptions of Newcastle English, providing the theoretical and empirical background for examining how Geordie speakers interact with and evaluate voice technologies.

\subsection{Sociolinguistic Biases in ASR}
\label{sec2.1}

Research on speech technologies and, more specifically, Automatic Speech Recognition (ASR) has consistently shown that technological performance such as recognition accuracy varies across social groups. Saying that this performance disparity is solely due to technical limitations is an oversimplification \citep{serditova25_interspeech}. Sociolinguistic research has drawn attention to the ways in which AI technologies can reproduce existing linguistic ideologies and hierarchies, while also introducing new forms of linguistic authority and inequality \citep{kelly2022sociolinguistics, erdocia2026language}. Previous work has primarily examined biases associated with gender, age, ethnoracial affiliation, and dialect, with less attention having been paid to socioeconomic status (SES) \citep{markl2022language, dichristofano2022global}. From a sociolinguistic perspective, however, SES is closely associated with patterns of pronunciation, lexical choice, and speech style \citep{labov1986social, guy1988language, milroy1993mechanisms, rickford1996regional}, raising the possibility that ASR systems may systematically perform differently across socioeconomic groups   \citep [cf.][]{serditova2026sociolinguistic}.

Evidence for demographic bias is mixed. Several studies report higher error rates for male speakers than female speakers \citep[however, see also][]{garnerin2021investigating, harris2024modeling}, while age-related effects tend to be weaker and less consistent, although children and older adults are often recognised less accurately \citep{liu2022towards, feng2024towards}. More robust evidence exists for ethnoracial disparities. Black speakers in the United States, particularly speakers of African American English (AAE), experience substantially higher word error rates than White speakers \citep{koenecke2020racial, liu2022towards, jahan2025unveiling}. These disparities have been linked to linguistic features characteristic of AAE, including phonetic, phonological, prosodic, and morphosyntactic variation \citep{lai2023exploring, Johnson2024, heuser24_interspeech, martin2020understanding, MojaradTang_2026_Probing_Interspeech}.

Dialectal variation has emerged as one of the most consistent sources of ASR bias. Across a range of English varieties, systems perform well on dialects that are well represented in training data but substantially worse on underrepresented regional and social varieties \citep{sanabria2023edinburgh, wassink2022uneven, lai2025dialect, markl2022language, serditova25_interspeech, serditova2026sociolinguistic}. Although expanding and adapting training data can improve recognition \citep{halpern2022low, sanabria2023edinburgh, torgbi2025adapting}, performance disparities often persist \citep{liu2022towards}, suggesting that ASR errors are rooted not only in data scarcity but also in systematic linguistic variation \citep{MojaradTang_2026_Probing_Interspeech, serditova2026sociolinguistic, scott2026sociophonetic}. Sociolinguistic research can contribute to understanding and addressing ASR disparities by identifying the social and linguistic sources of variation that speech technologies must accommodate \citep{serditova2026sounds}.

\subsection{Accommodation in Human--Computer Interaction}
\label{sec2.2}

Alongside questions of recognition accuracy, recent research has examined how speakers adapt their speech when interacting with voice-based technologies. Experimental studies have shown that speakers modify multiple aspects of their speech in response to perceived intelligibility problems, including speaking more slowly and loudly and articulating more carefully \citep{CohnZellou2021}. Similar patterns of phonetic and prosodic alignment have also been observed in routine interactions with voice assistants, suggesting that accommodation in human--computer interaction follows similar principles to human--human communication \citep{zellou2021age, ostrand2021its, gessinger2021phonetic}. Accommodation has also been observed in text-based human--AI interaction, with \citet{chen2026accommodates} finding that users gradually converged towards an LLM's function-word usage across conversational turns. 

This continuity between human--human and human--computer interaction is consistent with the Computers Are Social Actors (CASA) framework, which proposes that people respond socially to computers and may apply interactional norms derived from human communication even while recognising that they are interacting with a machine \citep{nass1994computers}. More recent extensions of CASA, however, suggest that repeated experience with technologies may lead users to develop distinct human--media interaction \citep{gambino2020building}.

Accommodation to voice technologies can also be understood through Communication Accommodation Theory (CAT). CAT proposes that speakers strategically modify aspects of their speech in response to the communicative context and their interlocutor, adjusting features such as pronunciation, dialect, vocabulary, and level of formality in order to achieve particular social or communicative goals \citep{giles1991language, giles2023communication}. These adjustments may take the form of convergence, i.e., speakers align their speech with that of their interlocutor, or divergence, i.e., speakers emphasise linguistic differences. They are well documented in collaborative dialogue and have been found to link to task success and interpersonal rapport \citep{hirschberg2008high, friedberg2012lexical, mitchell2012strangers, TianGriffithPriceBoyerTang_2024_LinguisticAlignment_LAS}.

Sociolinguistic research has long demonstrated that speakers do not produce a fixed version of their dialect. The use of regional features is influenced by socioeconomic background and situational factors, with speakers often reducing dialectal features in more formal settings and increasing their use in informal or emotionally engaged interactions, and also depending on the interlocutor and audience \citep{labov1966social, labov1972sociolinguistic, trudgill1974social, bell1984language, eckert2000linguistic}.

Interactions with voice assistants present a distinctive form of accommodation because the interlocutor is not another human but an automated system. Rather than adapting to a conversational partner's linguistic behaviour, users accommodate to their perception of what the technology is capable of recognising (i.e., perceived intelligibility). This may involve speaking more slowly or carefully, suppressing regional accents or avoiding local vocabulary in favour of forms perceived as more mainstream. While linguistic accommodation has been extensively studied in human interaction, comparatively little is known about how speakers of regional dialects adapt their speech to voice technologies.

\subsection{Geordie Accent Perceptions}
\label{sec2.3}

Newcastle English is a salient regional variety with strong associations with place and local identity. Perceptual dialectology research shows that speakers in North East England are sensitive to linguistic distinctions within the region and attach considerable social and geographical significance to local varieties \citep{pearce2009perceptual}. From a machine-perception perspective, a machine-learning-based classification of five Northern English accents (Leeds, Liverpool, Manchester, Newcastle upon Tyne, and Sheffield), based on first- and second-formant measurements across their full vowel systems, found considerable dialect levelling, particularly among Manchester, Leeds, and Sheffield, while Liverpool and Newcastle remained relatively distinct \citep{strycharczuk2020general}. More recent perceptual work suggests that listeners identify Newcastle voices more successfully than several other Northern English varieties, likely due to salient local features like glottalised /t/ \citep{montgomery2025feature}. Additionally, \citet{shaw2023revealing} demonstrate that Newcastle English has an accent-specific perceptual vowel structure and its perception depends on listeners' accent backgrounds (see also \citealt{best2015newcastle}). 

This distinctiveness is accompanied by complex social evaluations. Like other non-standard British varieties, Geordie accent perceptions are impacted by an existing hierarchy of prestige in which standard varieties such as SSBE tend to attract higher status evaluations, while regional varieties are frequently associated with lower socioeconomic status \citep{paver2025stereotyped}. In \citet{paver2025stereotyped}'s accent-perception study, the Newcastle voice was among the lowest rated for status and was perceived as particularly working-class and less confident.

However, there are positive attitudes towards the Geordie accent as well: \citet{coupland2007ideologised} found Newcastle English to be perceived as relatively pleasant, ranking considerably higher for social attractiveness than for prestige. Perceptions also differ between insiders and outsiders: \citet{burbano2008language} shows that Sunderland speakers strongly resist identification with Geordie speech, using perceived linguistic differences between the two areas as markers of distinct local identities. \citet{pearce2012folk} surveyed 165 local respondents about their perceptions of speech in Newcastle and Sunderland. He found that respondents from Tyne and Wear perceived Newcastle speech as broad, fast, melodic or ``sing-song'', and relatively harsh. Finally, representations of the North East in popular media can also play an important role in how the accent is perceived by those with limited direct experience of the region \citep{pearce2009perceptual, braber2015language}.

Such evaluations have real social consequences because speakers' competence, credibility, and behaviour might be stereotyped based on accent alone. Newcastle English provides a particularly useful case for examining regional voice AI: it is both a recognisably local and socially meaningful dialect and one that remains subject to social stigma.

\section{Method}
\label{sec3}

This section describes the study design and interview procedure (Section~\ref{sec3.1}), followed by the coding of the interview data and the statistical analysis of reported accommodation strategies (Section~\ref{sec3.2}).

\subsection{Data Collection}
\label{sec3.1}

\begin{figure}[H]
    \centering
    \makebox[\textwidth][c]{%
        \includegraphics[width=1\textwidth]{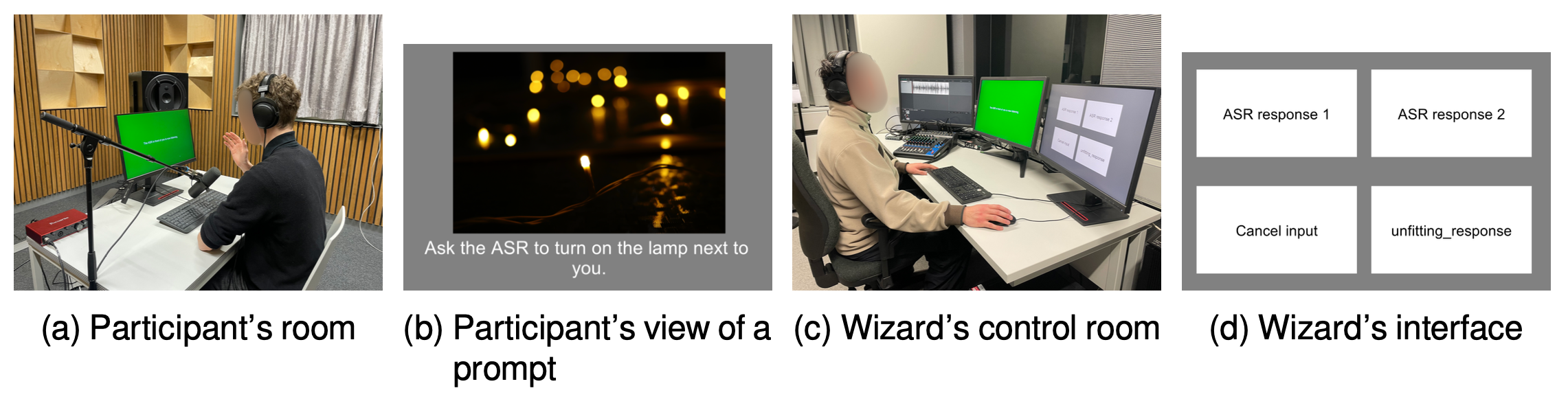}%
    }
    \caption{An illustration of the Wizard-of-Oz Setup. Left (a,b): participant, Right (c,d): wizard}
    \label{fig:experimental_setup}
\end{figure}

Data were collected in Newcastle upon Tyne using a Wizard-of-Oz experimental design, in which participants interacted with a simulated VA (see Figure~\ref{fig:experimental_setup} for the experimental setup). Depending on the experimental condition, the VA used either a Geordie or an SSBE voice; the system responses were pre-recorded by a voice actor who is bidialectal (Newcastle English and SSBE) and the interaction was controlled by the first author. Participants responded to guided prompts designed to elicit naturalistic speech while maintaining control over the linguistic features and interactional contexts targeted by the study.\footnote{See \url{https://doi.org/10.5281/zenodo.23685953} for the list of prompts.} Participants were balanced by gender and age and the type of system they interacted with (see ~\ref{app:participants} for the distribution table).

Following the interaction, participants completed a semi-structured interview designed to elicit reflections on their experience with the system, as well as their prior experiences with VAs, language use, and attitudes towards regional accents. The interview also collected relevant sociolinguistic background information, including age, gender, education, socioeconomic background, and regional orientation (see ~\ref{app:interview} for interview protocol).

The analyses reported in this paper draw on the semi-structured interview data rather than participants' speech produced during the experimental interaction itself.

\subsection{Interview coding and qualitative analysis}
\label{sec3.2}

The semi-structured interviews were coded in a spreadsheet organised around the interview questions (see ~\ref{app:participants}) and the topics relevant to the research questions. The coding was primarily descriptive rather than interpretive. For each participant, relevant excerpts from the interview transcript were copied verbatim into the corresponding categories in the spreadsheet. These included participants' descriptions of the VA voice they interacted with in the experiment before the post-hoc interview, evaluations of regional and mainstream-accented voices in both humans and machines, attitudes towards Newcastle English more generally, contexts in which they reported modifying their accent in everyday interactions, previous experiences of recognition difficulties with VAs they have at home, such as Siri or Alexa, and strategies used when a VA failed to understand them.

The coded interview data were used for both descriptive and qualitative analysis. Where responses could be grouped on the basis of explicitly stated similarities, recurring responses were counted to show how frequently they occurred across participants. For example, descriptions of the experimental VA voices were grouped under recurring descriptors such as \textit{familiar}, \textit{friendly}, \textit{clear}, or \textit{professional}. Participants could contribute to more than one category where their response contained multiple relevant descriptions or experiences. Counts therefore represent the number of participants who explicitly mentioned a particular response rather than mutually exclusive classifications.

The verbatim excerpts retained during coding were then used to examine how participants explained and contextualised these responses. The qualitative analysis focused on recurring patterns relevant to the research questions, such as associations between regional speech and familiarity or authority, perceptions of accent stigma, and participants' explanations of why they modified their speech. Quotations presented in the Results were selected from these coded excerpts to illustrate these patterns while preserving participants' own wording and the context in which the responses were expressed.

All coding was conducted by the first author. Because the procedure primarily involved recording and grouping explicitly stated responses while retaining the corresponding verbatim excerpts, rather than assigning participants' responses to abstract interpretive categories, intercoder reliability was not calculated. For variables subsequently quantified, a response was counted only when the relevant description, behaviour, or strategy was explicitly reported by the participant; it was not inferred by the coder.

\subsection{Quantitative analysis of accommodation strategies}
\label{sec3.3}

Participants' reported responses to recognition difficulties were grouped into six recurring accommodation strategies: enunciating, rephrasing, avoiding slang, speaking more slowly, standardising their speech or accent, and speaking louder. For the quantitative analysis of these strategies, mixed-effects binomial logistic regression models with a logit link were fitted using \texttt{glmer()} from the \texttt{lme4} package \citep{lme2015} in R \citep{r}.

Six strategies were coded as binary variables indicating whether each participant reported using the strategy when interacting with a VA: \textit{enunciating}, \textit{rephrasing}, \textit{avoiding slang}, \textit{speaking more slowly}, \textit{standardising their speech or accent}, and \textit{speaking louder} ($0 =$ not reported, $1 =$ reported). One participant (P001) did not complete this part of the study and was excluded, resulting in an analytical sample of $N = 57$. Because participants could report more than one strategy, the six responses were treated as repeated binary observations rather than as mutually exclusive categories. Each participant contributed one binary observation for each of the six accommodation strategies.

Categorical participant variables were coded as factors. Age, originally recorded in four groups (18--29, 30--44, 45--59, and 60+), was converted to approximate group midpoints (23.5, 37, 52, and 67, respectively) and entered into the model as a continuous predictor.
Gender\footnote{Gender was self-reported.} (man, woman) and regional orientation (local vs.\ external) were treated as categorical predictors and sum-coded ($-1$, $+1$). Education responses were initially heterogeneous and included several sparsely represented categories; these were therefore recoded into a binary distinction between participants with and without a university degree. Education information was unavailable for seven participants.

Participant was included as a random intercept to account for the non-independence of the six strategy observations contributed by each participant.  Models\footnote{An initial model included strategy type, age, gender, education, and regional orientation as fixed effects, with participant as a random intercept. Because inclusion of education reduced the sample from 57 to 50 participants due to missing data, an additional model excluding education was fitted to the complete accommodation dataset. The substantive pattern of results was unchanged, and the model without education was therefore used for the primary analysis so that all 57 participants (342 strategy-level observations) could be retained.} were fitted using the \texttt{bobyqa} optimiser with the maximum number of function evaluations increased to 200,000 to ensure convergence. A likelihood-ratio test was then used to compare this model with a more parsimonious model containing strategy type only, in addition to the participant random intercept. The demographic predictors did not collectively improve model fit, $\chi^2(3) = 2.52$, $p = .471$, and the reduced model also had a lower AIC (304.20 vs.\ 307.68). The strategy-only mixed-effects model was therefore retained, dropping age, gender, and regional orientation.

To test whether the likelihood of reporting accommodation differed across strategy types, the strategy-only model was compared with an intercept-only mixed-effects model containing the participant random intercept but no strategy effect. Inclusion of strategy type significantly improved model fit, $\chi^2(5) = 28.04$, $p < .001$, indicating that the six accommodation strategies differed in their probability of being reported.

Following the significant omnibus effect of strategy, estimated marginal means were calculated using the \texttt{emmeans} package \citep{emmeans}. Model-estimated probabilities and 95\% confidence intervals were obtained on the response scale for each strategy. Pairwise comparisons between strategies were conducted on the log-odds scale and expressed as odds ratios, with Tukey adjustment applied across the full family of pairwise comparisons to control for multiple testing. Model-estimated probabilities and their 95\% confidence intervals were also used for graphical presentation of the results.

\section{Results}
\label{sec4}

This section presents the results of the semi-structured interviews. Section~\ref{sec4.1} first examines participants' general perceptions of Newcastle English, establishing the social meanings and stereotypes associated with the accent. Section~\ref{sec4.2} then considers when participants modify their accent in everyday interactions, providing a sociolinguistic baseline for understanding accommodation to voice technologies. Section~\ref{sec4.3} turns to participants' open-ended characterisations of the Geordie and SSBE VA voices they encountered during the experiment, before Section~\ref{sec4.4} examines the social implications of a Geordie-speaking VA. Finally, Section~\ref{sec4.5} examines the strategies participants reported using to make themselves better understood by voice AI and considers how these relate to the accommodation practices described in everyday interactions.


\subsection{Newcastle English Perceptions}
\label{sec4.1}

Participants were first asked about their perceptions of the Geordie accent more generally, providing a sociolinguistic context for the subsequent analysis of the Geordie VA. Many expressed frustration with the stereotype that the Geordie accent sounds uneducated or less intelligent:

\begin{quote}
``I think southerners probably listen to our accent and think we're stupid and thick and ignorant [\ldots] aren't as cultured and intelligent as perhaps they perceive themselves to be [\ldots] I have been aware of occasions where, um, people have heard the accent and have just turned their noses up and walked away.'' --- P027 (M, 45--59)
\end{quote}


\begin{quote}
``[I]t can be seen as thick [\ldots] lower class, potentially, which, I don't think that's right. I think potentially in the media we have a bit of a bad reputation. [\ldots] You can't help where you're born and your accent is your accent.'' --- P041 (W, 30--44)
\end{quote}

Several university students who are women reported having personally experienced such accent-based judgements. For example, P014 (W, 18--29), a mechanical engineering student, described how people from outside the region were sometimes surprised to learn that she attended university, suggesting that assumptions about her educational background were made on the basis of her speech. Another informant expressed her frustration with how unintelligent the Geordie accent sounds on TV and that she herself wants to sound more professional because of that. And yet another participant shared that she was encouraged to change her accent at school in order to be more employable (it is notable that this informant is young, meaning that these situations occur in the modern day and are not just remnants of the past).

Overwhelmingly positive perceptions were expressed as well, albeit less commonly. Those participants called the accent reassuring, touching, positive, happy, and friendly.

\begin{quote}
``I think Newcastle are deemed as one of the friendliest because of, partly because of the accent.'' --- P045 (W, 30--44)
\end{quote}

In many cases, perceptions were mixed. Participants pointed out that the Geordie dialect might be perceived as not as intelligent sounding as opposed to the Queen's English or London-centric English, but also called it warm and welcoming. A participant from the age group 18--29 who is a woman noted that outsiders think the Geordie accent is either really fun, happy and positive or a ``poor accent'', signifying that one does not have class or is not financially well-off. She and numerous other respondents noted that growing up, parents would tell their kids to speak properly, otherwise they would not be able to go to university. She also noted that people like to copy and imitate the accent, and that in the media, Geordies are portrayed as happy presenters.

One participant summarised the contrasting perceptions:

\begin{quote}
``Well, I've heard two things. Um, you, you hear conflicting information. So one, the positive one is that people tend to think that Geordie accent is very friendly. We've got the reputation of being very friendly. So I think a lot of people like it, they like the Geordie accent and feel that they are friendly, but then there's also a negative perception and I think people think that if you speak with a Geordie accent, you're not very clever. So there's that, um, you know, that person can't be a lawyer or a doctor or with a strong Geordie accent. It's sort of associated with say, working class manual labour and miners, ship builders, that sort of thing. Um, so I think, yeah, there's that negative perception of, we're friendly and open and nice, but we're not professionals or very intelligent.'' --- P022 (W, 45--59)
\end{quote}

The accent was also linked to industry, working class, and labour. Study contributors pointed out that these perceptions are outdated since the shipyard industry is no longer there.
In some cases, these outdated perceptions remain, in other cases, they have been substituted. Numerous participants highlighted that the Geordie accent and Geordies themselves are now associated with drinking and partying:

\begin{quote}
``People think if you're from Newcastle, you love to go out drinking.'' --- P006 (W, 18--29)
\end{quote}

Given the role of media representation in how regional accents are perceived \citep{pearce2009perceptual,braber2015language}, participants were also asked about the representation of Geordies in mainstream media. They generally perceived the accent as underrepresented, particularly in contexts associated with authority and formality, such as news broadcasting. Only two participants (M, 18--29, and W, 45--59) considered its representation broadly adequate, with one pointing to the increasing presence of regional voices across television, film, radio, and news. Several participants nevertheless felt that media representation had improved over time, frequently crediting prominent local figures such as \textit{Ant \& Dec} with increasing the visibility and positive recognition of the accent. At the same time, participants distinguished between greater visibility and the kind of representation available: programmes such as \textit{Geordie Shore} were repeatedly cited as negative or stereotypical portrayals.



Participants recognised that social judgements often involve attributing social characteristics to speakers on the basis of accent alone. There was an awareness that one's accent is largely a consequence of where and how one grows up and does not inherently imply particular attitudes, abilities, or personality traits. P051 (M, 18--29), for example, rejected assumptions that speakers of non-RP varieties are necessarily ``brutish'' or prejudiced, emphasising that he speaks his dialect simply because he was born in the region rather than because it reflects any particular set of attitudes.





\subsection{When Do Geordies Feel the Need to Modify Their Accent?}
\label{sec4.2}

This part of the interview explored (1) the contexts in which Geordie speakers modify their accent in everyday interactions outside the use of speech technologies, and (2) participants' metalinguistic awareness of such accommodation. The latter is particularly relevant given previous research demonstrating high levels of awareness of locally distinctive linguistic features among speakers of Newcastle English \citep{beal2009enregisterment, pearce2017linguistic}. Establishing these patterns provides an important point of comparison for examining whether and how speakers draw on similar forms of accommodation when interacting with voice technologies. Professional contexts were the most frequently reported setting for accent modification, mentioned by 18 participants (31.6\%), followed by travel (14; 24.6\%) and interactions with family and friends (12; 21.1\%).

Motivations for workplace accommodation varied: some participants described modifying their speech to facilitate communication with non-local colleagues, while others associated a less regionally marked accent with sounding more ``proper'', ``professional'', or ``polite''. Several referred to this style as their ``telephone voice'', although the term was not restricted to telephone conversations and could describe a more mainstream speech style used across professional or formal situations. Discussions of the ``telephone voice'' also reproduced some of the language attitudes identified above, particularly the perception that a broader Geordie accent may be evaluated as less intelligent.

Accent modification did not, however, invariably involve reducing regional features. 12 participants (21.1\%) reported using a broader or more recognisably Geordie accent with family and friends.


\begin{quote}
``I work in an office and yeah, Geordie accent gets toned down at times. [\ldots] It's definitely softer when I'm at work [\ldots] But if I've been around me sister a lot more [\ldots] I probably, the more angry I get, I probably do. I, I probably do get more Geordie if I, I'm trying to think if I'm shouting at the kids and tell them off.'' --- P053 (W, 60+)
\end{quote}



Travel was another common context for accommodation, mentioned by 14 participants (24.6\%). Participants described both reducing regionally marked features and slowing their speech when travelling, particularly when visiting other parts of the UK or interacting with people who might be less familiar with the accent.



Many of these accounts can be understood as instances of convergence, in which speakers adjusted their speech towards that of their interlocutors or towards a style they perceived as more appropriate to the interaction. Participants frequently described these shifts as automatic or subconscious rather than deliberate. Regional features could become more salient when speaking to other Geordies, while interactions with non-local or more formal interlocutors could result in less dialectal speech.


\begin{quote}
``I know that I talk more Geordie when I'm talking to somebody Geordie. [\ldots] But when you're talking to somebody professional or from down south [\ldots] you do telephone voice.'' --- P050 (W, 60+)
\end{quote}


There were also two cases of accent divergence reported by participants, who claimed that their Geordie accent becomes stronger when they are around non-Geordies:

\begin{quote}
``I think when I'm around people who don't speak it, I almost, it just comes out more.'' --- P019 (W, 18--29)
\end{quote}

\begin{quote}
``I think when I've gone abroad it's gotten stronger a couple of times.'' --- P058 (M, 18--29)
\end{quote}

Finally, four participants (7\%) expressed their strong unwillingness to change their accent under any circumstances, framing their speech as an important part of regional identity rather than something that should be adjusted according to context:

\begin{quote}
``Why would you need to change the way you speak? Maybe to sound more intelligent?'' --- P024 (W, 60+)
\end{quote}

\begin{quote}
``Geordies never [change the way they speak]. If you were born Geordie, you'd be so proud of where you come from.'' --- P026 (M, 45--59)
\end{quote}

Importantly, resistance to accommodation did not imply a lack of metalinguistic awareness. Even participants who reported never modifying their speech recognised that their accent was socially salient; as P030 (W, 45--59) put it, ``No but I know your accent is noticed''. Thus, participants' accounts reveal awareness of both their own linguistic accommodation and the social meanings attached to their accent, even where they did not report consciously modifying it.

\subsection{Characterisations of the Experimental VA Voices}
\label{sec4.3}

Participants were asked to describe the VA voice they encountered during the experiment. Their responses revealed distinct characterisations of the Geordie and SSBE voices, particularly in relation to regionality, familiarity, warmth, clarity, and professionalism. The following sections consider each voice in turn.

\subsubsection{Characterisations of Geordie VA}
\label{sec4.3.1}

\begin{figure}[H]
    \centering
    \makebox[\textwidth][c]{%
        \includegraphics[width=1.1\textwidth]{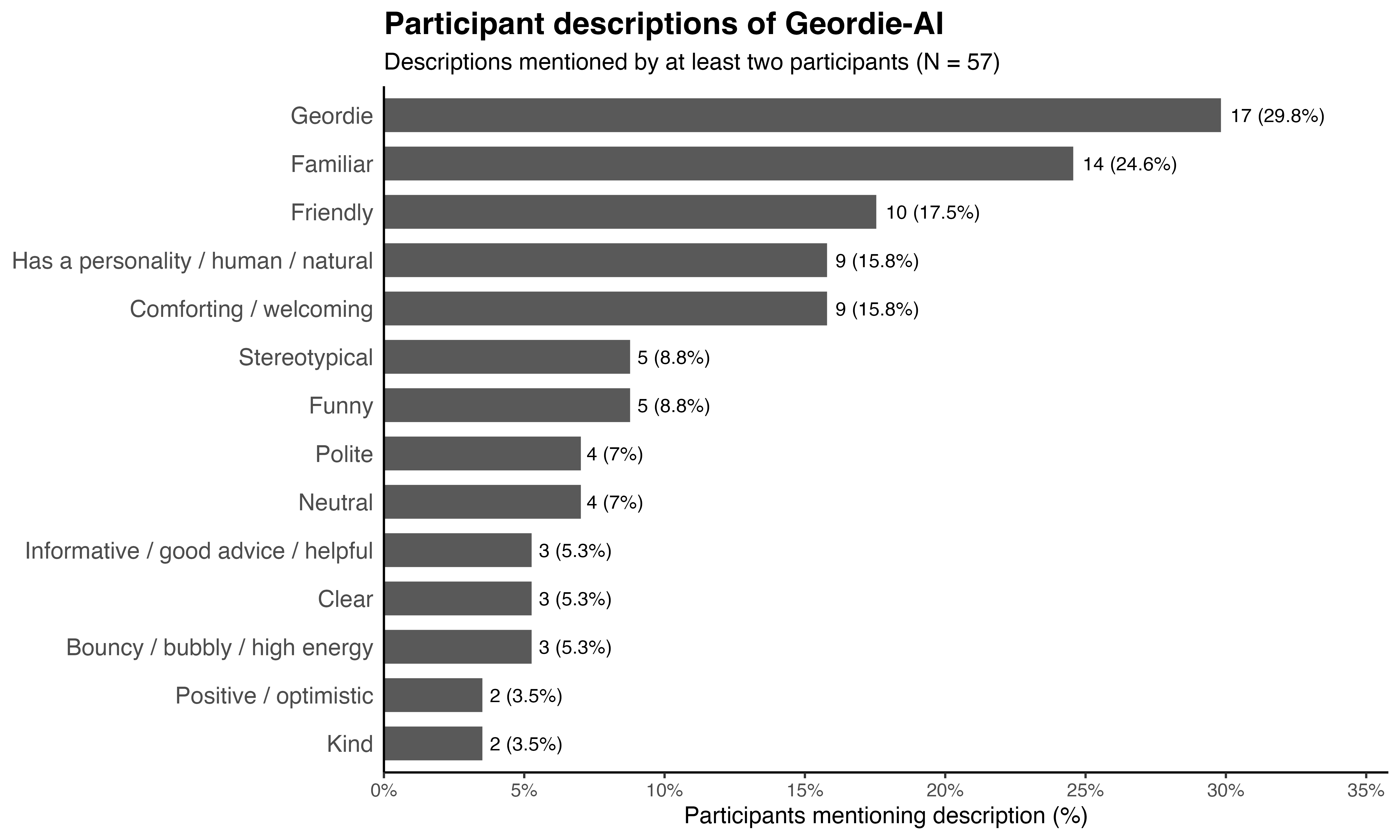}%
    }
    \caption{Participant descriptions of the Geordie VA.}
    \label{fig:geordie_va_feelings}
\end{figure}

As shown in Figure~\ref{fig:geordie_va_feelings}, descriptions of the Geordie-AI voice were predominantly associated with regional identity, familiarity, and interpersonal warmth. The most frequently mentioned characteristic was \textit{Geordie} (29.8\%), followed by \textit{familiar} (24.6\%), \textit{friendly} (17.5\%), \textit{human/natural or having a personality} (15.8\%), and \textit{comforting/welcoming} (15.8\%). Less frequent descriptions included \textit{stereotypical} and \textit{funny} (both 8.8\%), while attributes relating more directly to clarity, informativeness, and positive affect were mentioned by smaller proportions of participants. Overall, the responses suggest that participants primarily characterised the voice in terms of its regional and socially engaging qualities, rather than its functional characteristics. Thus, most descriptions were positive, except for the stereotypical ones.

Those who described the voice as familiar often remarked that it resembled the speech of someone they knew, such as family members, or even their own way of speaking. It is not a surprise that this made the Geordie system sound relatable, warm, or trustworthy.

\begin{quote}
``It sounded more like my family members, like my mom. So it almost felt more comforting to talk to.'' --- P014 (W, 18--29)
\end{quote}

\begin{quote}
``So the voice was very familiar. I, it was strange though. 'cause even some of the nuances of words sound like how I would say them.'' --- P051 (M, 18--29)
\end{quote}

\begin{quote}
``It was quite funny to hear the [\ldots] Geordie accent. [\ldots] It felt like it was almost mirroring what I was saying, how I was pronouncing things.'' --- P052 (M, 45--59)
\end{quote}

Participants admitted that because the voice sounded friendly, warm, and welcoming, the interaction felt more like a real conversation as opposed to VAs they are used to hearing, which were described as robotic. However, one participant (W, 18--29), who also referred to the voice as friendly, positive, and optimistic, admitted that for that reason it does not sound professional. This raises the question whether the same voice qualities associated with familiarity and warmness are compatible with professionalism and credibility in a VA.

Some participants also described the voice as stereotypically Geordie:

\begin{quote}
``I think it was good, but my sense is that it was a voice trying to be Geordie as opposed to a Geordie voice.'' --- P039 (W, 45--59)
\end{quote}

\begin{quote}
``It felt like a person because of the intonation of the accent. [...] A little too localised. And I feel that sometimes in the tone, there is a kind of condensation, like the, you know, you're being talked down to. Um, and it's almost like playing on the accent more. It felt too stereotypical.'' --- P038 (W, 60+)
\end{quote}

Although such evaluations were relatively uncommon (5 out of the 35 participants (14.3\%) who interacted with the Geordie system mentioned it), they point to a potential conflict between regional authenticity and stereotypical representation. For some participants, the presence of recognisably Geordie features did not necessarily make the voice sound authentically local but rather like it performs Geordieness. Increasing the perceptual salience of regional features may reach a point at which regional representation is interpreted as exaggeration or caricature. Notably, these perceptions arose despite the VA voice having been recorded by a native Newcastle English speaker, demonstrating that the authenticity of the speaker does not directly translate into perceived authenticity of the voice. Listeners' evaluations appear to depend on how particular accent features are interpreted within the context of a synthetic or AI-mediated interaction.

\subsubsection{Characterisations of SSBE VA}
\label{sec4.3.2}

\begin{figure}[H]
    \centering
    \makebox[\textwidth][c]{%
        \includegraphics[width=1.1\textwidth]{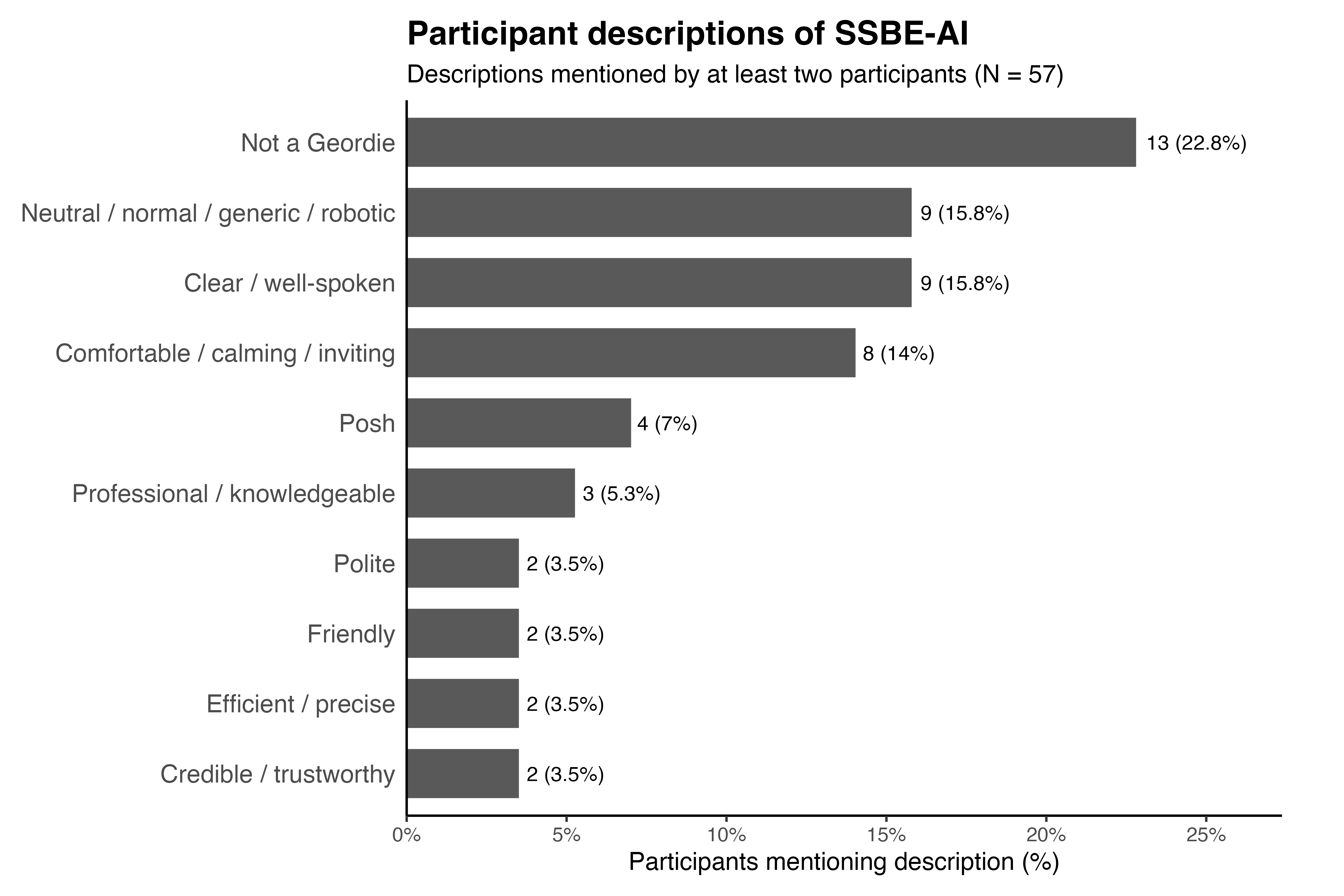}%
    }
    \caption{Participant descriptions of the SSBE VA.}
    \label{fig:ssbe_va_feelings}
\end{figure}

In contrast to attitudes towards the Geordie VA, descriptions of the SSBE VA voice focused less on familiarity and regional identity and more on its non-local character, neutrality, and clarity. The most frequent description was \textit{not a Geordie} (22.8\%). This response should be interpreted in light of the experimental framing: participants had been told that they were helping to test a Geordie VA, making the absence of a Geordie accent particularly salient to those assigned to the SSBE condition. Other frequent descriptions included \textit{neutral/normal/generic/robotic} and \textit{clear/well-spoken} (both 15.8\%). The voice was also described as \textit{comfortable/calming/inviting} by 14\% of participants. Less frequent descriptions included \textit{posh} (7\%) and \textit{professional/knowledgeable} (5.3\%), while attributes such as friendliness, politeness, efficiency, and credibility were mentioned by relatively few participants (Figure~\ref{fig:ssbe_va_feelings}).

Further comments by participants reinforced the lack of strong feelings towards the SSBE system.

\begin{quote}
``Uh, well this voice is, uh, very neutral. Um, there's nothing offensive about it, but there's nothing particularly warm and attractive about it either.'' --- P022 (W, 45--59)
\end{quote}

\begin{quote}
``There wasn't the connection or there wasn't as much of a deep connection than I think would've been with the Geordie voice.'' --- P044 (M, 60+)
\end{quote}

Overall, participants characterised the SSBE voice as relatively neutral and clear, but less personally familiar than the Geordie voice. Some also associated its neutrality with professionalism.


\subsection{Geordie VA Perceptions}
\label{sec4.4}

Following the contrasting characterisations of the Geordie and SSBE VA voices, participants were asked more explicitly about the social implications of these differences, including whether a Geordie-speaking VA might be perceived as less authoritative than one using a more mainstream dialect such as SSBE. Most participants (n=54, 94.7\%) believed that this could indeed be the case. Their explanations frequently drew on the social stigma attached to Northern English accents: participants suggested that advice delivered by a Geordie-speaking system might be taken less seriously, perceived as less credible, or judged as less reliable than the same information delivered in a more mainstream accent. These concerns were particularly salient in higher-stakes contexts, such as receiving medical advice. At the same time, other participants did not necessarily view regionality as a disadvantage in such contexts. Some emphasised the familiarity and reassurance afforded by a local voice, with one stating that they would feel more comfortable ``taking medical advice from someone who mirrors [their own] accent'' (P025, W, 18--29).

For some participants, regionality also contributed to perceived naturalness and intelligibility. P058 (M, 18--29), for example, found the Geordie voice ``easier to understand [\ldots] and [it] sounds more natural to listen to rather than like received pronunciation like Southern English''.

Participants demonstrated considerable awareness that these evaluations were socially learned rather than inherent properties of the accent or of the information provided by the system. Several explicitly described the possibility of internalised accent bias:

\begin{quote}
``I think there might be a little bit of inner bias in there somewhere, you know, subconsciously.'' --- P027 (M, 45--59)
\end{quote}

\begin{quote}
``I think listening to like a Geordie accent on this type of thing, I think yeah, definitely has that kind of less authoritative, um, manner in the way that the responses come.'' --- P036 (W, 18--29)
\end{quote}

This conflict was articulated particularly clearly by P022, who recognised a discrepancy between what users rationally know about AI and the social evaluations that a human-like voice may nevertheless evoke:

\begin{quote}
``I think anybody who knows how AI works shouldn't, or, or would, wouldn't or shouldn't think that the accent makes a difference because you know that it's scraping information from the Internet and it doesn't really, and it's not a person anyway, so it's, so it shouldn't, but the question is, does it subconsciously, because you're almost seeing it as a person and those values are being transferred onto the, oh gosh, I don't know. I, I just don't think I can answer that.'' --- P022 (W, 45--59)
\end{quote}

The human-like social persona associated with the regional voice could consequently influence perceptions of what kinds of advice it was suited to provide. For example, P055 compared the Geordie system to an auntie or someone who gives informal advice:

\begin{quote}
``I could see that definitely when it comes to kind of medical advice and stuff, because like even how I said I see it kind of like a, an auntie that you see every now and then. Um, it's kind of like the advice you would get in school, maybe [\ldots] Um, I feel like that is how I could be perceived.'' --- P055 (W, 18--29)
\end{quote}

Participant P037 (W, 18--29) similarly questioned whether a Geordie voice could convey the same degree of authority as a more mainstream-accented voice. She suggested that because the Geordie accent is frequently made fun of and has not traditionally been associated with a ``voice of authority'', its use by a VA or an AI system might initially be perceived as humorous rather than authoritative. A similar concern was expressed by P038 (W, 60+), who acknowledged that a regional voice might make some users feel more comfortable, but suggested that others could interpret the speaker as less capable or qualified to perform more professional tasks, despite knowing that the system was AI. Thus, existing associations between accent and professionalism or authority may carry over into interactions with voice AI.

Participants did not necessarily expect such evaluations to be shared equally across listeners. Several located accent stigma geographically, suggesting that a Geordie voice would be evaluated more positively within the North East than elsewhere in England. P057, for instance, anticipated bias but distinguished between local and non-local listeners: ``[even among other Geordies?] Oh, I don't think among other Geordies, but I think maybe other, like down south'' (P057, M, 18--29). London emerged particularly frequently as a point of contrast:

\begin{quote}
``Probably people from London would stigmatise it but here it's normal, so no.'' --- P028 (M, 18--29)
\end{quote}

\begin{quote}
``I feel like it would probably depend on who it is as well. Like, people up here would probably like it more, but people in London or something might not think of it as much.'' --- P035 (M, 18--29)
\end{quote}

Interestingly, the previously discussed \citet{shaw2023revealing} found Newcastle and London to be the least perceptually similar of the five English varieties they examined (see Section~\ref{sec2.3}). While perceptual similarity does not necessarily correspond to social evaluation, this finding is notable given participants' repeated positioning of London as a locus of anticipated accent stigma.

Many participants also expressed a concern that a Geordie VA would simply not be accessible to those who were not born and raised in Newcastle:

\begin{quote}
``Being a Geordie, not everyone can understand a Geordie accent. There are people from all over the country.'' --- P024 (W, 60+)
\end{quote}

At the same time, some believed that it would be a positive thing if visitors or newcomers were exposed to the local accent more:

\begin{quote}
``[P]eople come to the city, um, they know what the accent's like and maybe they can be a bit more tuned to that. Um, and I think for people who are in the city, it's nice to hear like a local accent when you like, like you said, the train station or the buses or the metros, things like that.'' --- P041 (W, 30--44)
\end{quote}

\begin{quote}
``[S]ay if people who were travelling up here for like whatever reason, it gives them a little sense of, oh, but we're in Newcastle, we've got the accent around.'' --- P055 (W, 18--29)
\end{quote}

Participants also recognised that evaluations of VAs may be impacted by the interaction between accent, gender, and socially constructed expectations of different voices. In particular, participants associated female voices with qualities such as gentleness, helpfulness, and kindness, while male voices were sometimes perceived as potentially more forceful or authoritative:

\begin{quote}
``So one odd thing on that, I think a female one would be taken as more gentle and more sort of like, sort of like it's got less edge. Whereas if it was a man, I feel like there's a bit more of an aggression.'' --- P051 (M, 18--29)
\end{quote}

\begin{quote}
``[E]ven without knowing it, we've all got these internal biases. [\ldots] like there's a reason that they choose women's voices for these mostly, isn't it? Like Alexa and all these things have got women's voices because we're expecting them to help and to give us information. But there's a reason why it's usually a man who'd be the doctor or whatever.'' --- P049 (W, 45--59)
\end{quote}

P056 identified the challenge for VAs and voice AI as combining warmth and authoritativeness within the same voice:

\begin{quote}
``I think the, the, the tone of it could suggest that, um, maybe it might be different with a male voice. But maybe that's just like society as a whole. [\ldots] But maybe that's the challenge to get it kind but authoritative.'' --- P056 (W, 30--44)
\end{quote}

Thus, perceived authority cannot be attributed to accent alone. Evaluations of VAs and voice AI are connected to multiple socially meaningful characteristics of a voice, including regionality and gender.

In total, only three participants (5.3\%) explicitly rejected the suggestion that a Geordie-speaking VA might be evaluated less favourably than a system using a more mainstream accent. However, participants tended to locate the anticipated stigma outside the local community rather than among Geordie users themselves. Participants also recognised that such evaluations need not remain fixed: P034 (W, 60+) suggested that greater exposure to different accents could lead to greater acceptance. Increased representation of regional voices in technological contexts may therefore have the potential not only to reflect linguistic diversity, but also to normalise varieties that users are less accustomed to hearing in these domains.

At the same time, participants did not necessarily favour maximising the regional character of the system. When asked whether local lexical items should be incorporated into the VA's vocabulary, responses were generally skeptical. Participants worried that deliberate use of Geordie vocabulary could sound unusual or overly performative. P048 (M, 18--29), for example, felt that repeatedly inserting local forms could make the voice sound ``forced'' and consequently less natural. 

These responses also raise the question of how closely a VA or voice AI needs to resemble its users in the first place. As P039 (W, 45--59) observed, accepting that AI voices are not human may also mean accepting that ``it's okay if they don't sound like you''. Across the interviews, the appeal of the Geordie VA voice seemed to be less in simply reproducing as many local features as possible but in providing a recognisably regional voice that could evoke familiarity, belonging, and inclusion without appearing exaggerated or stereotypical.

\subsection{Strategies to Be Better Understood by Voice AI}
\label{sec4.5}

The communication accommodation described above was not limited to interactions with other speakers. Participants reported drawing on similar strategies when attempting to make themselves understood by a VA, although the particular adjustments they made varied considerably. In this context, accommodation was directed not towards another speaker, but towards participants' perceptions of what the technology was capable of recognising. This raises the question: what, specifically, do Geordie speakers change in their speech in order to be understood by the machine?

Descriptively, speaking with a more mainstream accent was the most commonly reported accommodation strategy, mentioned by 18 participants (31.6\%). This was followed by enunciating more clearly (n=16, 28.1\%) and speaking more slowly (n=14, 24.6\%). Rephrasing was reported by eight participants (14.0\%), while avoiding slang (n=5, 8.8\%) and speaking louder (n=3, 5.3\%) were less common. As participants could report multiple strategies, these categories were not mutually exclusive.

\begin{figure}[H]
    \centering
    \makebox[\textwidth][c]{%
        \includegraphics[width=1.2\textwidth]{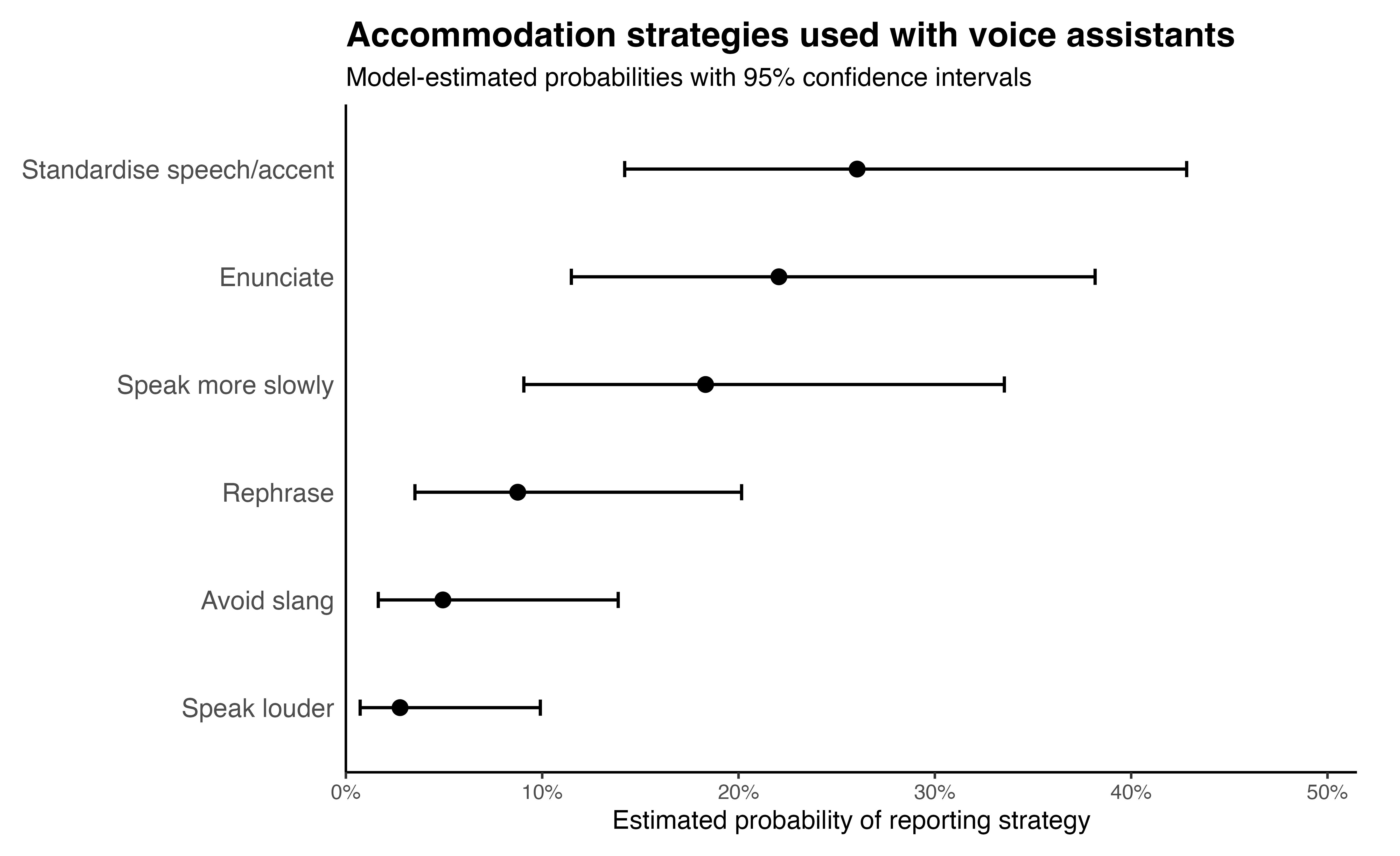}%
    }
    \caption{Accommodation strategies participants use with VAs.}
    \label{fig:accommodation_strategy_probabilities}
\end{figure}

To test whether these descriptive differences varied systematically by strategy type, we examined the strategy effect in the mixed-effects model. Including strategy type significantly improved model fit relative to the intercept-only model, $\chi^2(5) = 28.04$, $p < .001$. As shown in Figure~\ref{fig:accommodation_strategy_probabilities}, the model-estimated probability was highest for standardising speech or accent (26.0\%, 95\% CI [14.2, 42.8]), followed by enunciating more clearly (22.1\%, 95\% CI [11.5, 38.2]) and speaking more slowly (18.3\%, 95\% CI [9.1, 33.5]). Lower probabilities were estimated for rephrasing (8.8\%, 95\% CI [3.5, 20.2]), avoiding slang (4.9\%, 95\% CI [1.7, 13.9]), and speaking louder (2.8\%, 95\% CI [0.7, 9.9]).

Tukey-adjusted pairwise comparisons showed that speaking louder was significantly less likely to be reported than enunciating ($p = .017$), speaking more slowly ($p = .046$), and standardising speech or accent ($p = .006$). Avoiding slang was also significantly less likely to be reported than standardising speech or accent ($p = .021$). No other pairwise differences reached statistical significance after adjustment for multiple comparisons.

\subsubsection{Strategy 1: Standardise Speech/Accent}
\label{sec4.5.1}

Standardising speech or accent was the most frequently reported strategy. Participants described deliberately reducing features they perceived as recognisably Geordie and moving towards forms they believed voice AI would be more likely to recognise. The target of this accommodation was not necessarily a single clearly defined accent: participants described adopting a more American, Southern English, ``posh'', or otherwise more neutral way of speaking.

\begin{quote}
``Most of the time I have to put an American accent.'' --- P014 (W, 18--29)
\end{quote}

\begin{quote}
``Sometimes it's quicker to just do it yourself. I'd try to lower the accent a bit. Especially with a lot of them being American. I think it's quite difficult for them to understand the northern accent. So I think you have to kind of like southernise the accent a bit. A bit posher like, proper English.'' --- P028 (M, 18--29)
\end{quote}

Participants used several recurring labels to describe this accommodation. These included ``taking the Geordie off'', speaking ``proper English'', trying to ``sound more neutral'', making the voice ``posher'', and switching to a ``telephone voice''. Therefore, successful communication with voice AI implied movement away from dialectal speech and towards varieties perceived as more standard or technologically intelligible. Several participants mentioned specific linguistic adjustments, such as pronouncing their /t/s more clearly, referring to the avoidance of /t/-glottalisation. For some, this accommodation required conscious monitoring of speech; as one participant (M, 45--59) explained, speakers had to think about how they would vocalise an utterance before saying it, otherwise the technology might fail to recognise it.

Not all participants felt equally able to make these adjustments. For example, P031 expressed the frustration behind not being understood and not being able to change the way he speaks enough to ``please'' the system:

\begin{quote}
``It's just, just a little bit frustrating if I'm quite honest. Um, you know, you, it just actually cannot understand your word and there's no way you can change how you say it that they can understand you. So there's people within my family, like me children, me two sons who can say something who are probably not as Geordie as I am. And they can be understood with a Northeastern accent.'' --- P031 (M, 45--59)
\end{quote}

These accounts suggest that participants had developed a relatively explicit model of the kind of speech voice technologies are more likely to recognise \citep{serditova2026sociolinguistic, serditova25_interspeech}.

\subsubsection{Strategy 2: Enunciate}
\label{sec4.5.2}

The second most frequently reported strategy was enunciating more clearly, mentioned by 16 participants (28.1\%). Participants described making a conscious effort to articulate words more carefully when voice AI failed to recognise their usual speech. Unlike accent standardisation, this strategy did not necessarily involve adopting a different or less regional accent. In some accounts, clearer articulation occurred specifically in response to repeated recognition failures and the frustration they caused. As P003 (M, 18--29) explained in relation to Alexa:

\begin{quote}
``I lose my head at Alexa sometimes. I use Alexa a lot 'cause just to put music on and that. But like if I want the music changed or anything, it can get on my nerves or it doesn't stop straight away. I just have to speak more clearly.'' --- P003 (M, 18--29)
\end{quote}

Notably, the boundary between enunciation and accent standardisation was not always clear-cut, as careful articulation could involve suppressing locally characteristic pronunciations. This overlap also reflects the fact that participants often combined several accommodation strategies rather than treating them as distinct adjustments.

\subsubsection{Strategy 3: Speak More Slowly}
\label{sec4.5.3}

Speaking more slowly was reported by 14 participants (24.6\%). This strategy was frequently connected to participants' perceptions of the pace of Newcastle English, which they characterised as relatively fast. Slowing down therefore represented a relatively straightforward way of adapting their usual speech to what they perceived as the processing limitations of VAs. As P002 (W, 60+) explained, an instruction such as ``Alexa, turn the lights out now'' might not be recognised because she had ``spoke[n] too fast''.

Slowing down also frequently occurred alongside other forms of accommodation. P052 (M, 45--59), for example, described both slowing down and ``slightly modify[ing]'' his speech, explicitly comparing this behaviour to the way he accommodates when speaking to non-Geordies. This provides a particularly clear connection between accommodation in human--human and human--computer interaction: participants could draw on strategies already familiar from communication with non-local speakers and apply them to voice technologies.

However, accommodation did not guarantee successful recognition. P045 (W, 30--44) reported speaking ``really clear and slow'' when using speech-to-text software, yet still experiencing transcription errors. Thus, participants may undertake considerable communicative effort to accommodate the system while recognition problems nevertheless persist.

\subsubsection{Strategy 4: Rephrase/Repeat}
\label{sec4.5.4}

Repeating or rephrasing a request was reported by eight participants (14.0\%). Unlike the phonetic adjustments described above, this strategy involved the wording of the request: participants either repeated an utterance following a recognition failure or reformulated it in an attempt to make the intended request easier for the system to process.

Repetition also appeared to have a relatively clear limit. When repeated or reformulated requests continued to be unsuccessful, participants described giving up on the interaction altogether and turning the VA off. This suggests that accommodation to voice technologies involves not only adjustments to speech, but also decisions about how much communicative effort users are willing to invest before disengaging from the technology.

\subsubsection{Strategy 5: Avoid Slang and Local Lexical Items}
\label{sec4.5.5}

Avoiding slang and local lexical items was reported by five participants (8.8\%). Here, accommodation involved specific lexical choices, with participants replacing regional lexical items with forms they perceived as more likely to be recognised by VAs. P002 (W, 60+) described having to rephrase her ``slanguage'' in ``proper English'', while P004 (M, 30--44) similarly reported replacing dialect or slang with ``standard English or standardised English words''.

Lexical accommodation often overlapped with other strategies: changing a dialect word could involve rephrasing the request (2/5 participants mentioned it) while simultaneously modifying the strength of the accent (3/5 participants mentioned it). As with accent standardisation, participants had an expectation of what voice technologies are able to understand, with regional forms being perceived as less technologically intelligible.

\subsubsection{Strategy 6: Speak Louder}
\label{sec4.5.6}

Speaking louder was the least frequently reported accommodation strategy, mentioned by three participants (5.3\%). As with the other strategies, increased volume could occur alongside additional modifications; P022 (W, 45--59), for example, reported that she would ``speak more loudly'' and ``probably enunciate more'' when addressing a VA.

The relative rarity of this strategy is noteworthy given that increased loudness is a commonly documented response to communication difficulties with technological or otherwise challenging interlocutors \citep{cohn2022acoustic, CohnZellou2021, cohn2024african, zellou2024linguistic}. However, the present data suggest that participants generally interpreted recognition failures as problems of how they were speaking, rather than whether the system could hear them.

While most participants reported accommodating their speech to VAs, only five of the 57 participants (8.8\%) reported generally positive experiences with VAs and no substantial recognition difficulties. One attributed this to having a relatively neutral accent, while the others provided little explicit explanation. Notably, all five were middle-class speakers; three had an external place orientation (i.e., they felt less strongly connected to Newcastle and were more open to living elsewhere; see \citet{labov1963social, carmichael2025language, monka2020place, serditova2026sounds}), and one of the two locally oriented participants reported little connection to their neighbourhood. These characteristics may suggest that speakers whose linguistic practices are less strongly locally oriented experience fewer recognition difficulties. However, given the very small number of participants and the absence of direct evidence about their actual speech production, this interpretation should remain tentative: social class and place orientation cannot be assumed to correspond straightforwardly to a more mainstream accent or to explain their greater success with voice technologies.

Overall, participants rarely relied on a single accommodation strategy. Instead, their accounts described combinations and sequences of adjustments, often escalating when an initial attempt failed:

\begin{quote}
``You change your voice. I give it perhaps one or two goes speaking normally. And if that doesn't work, then I try, try and pronunciate very slowly like this and I may take my accent up or down a little bit to try and make it work. And then you throw against the wall.'' --- P027 (M, 45--59)
\end{quote}

\begin{quote}
``It doesn't pick up what you're saying at all. [\ldots] obviously there's like, lexical differences which like I think are also like important to recognise. [\ldots] I think it's more, it is just more going for like a standard and like articulating, like actively articulating and like breaking up the words. Especially like with speed as well. I feel like the Geordie accent is quite a fast like dialect and it just is just not, it just doesn't compute all those words.'' --- P037 (W, 18--29)
\end{quote}

Accommodation to voice technologies is best understood not as a set of isolated strategies, but as a fluid process of communicative adjustment. Participants appeared to diagnose what they believed was preventing recognition and modify multiple dimensions of their speech accordingly, sometimes moving progressively further away from their habitual speaking style.

\section{Discussion}
\label{sec5}

The findings demonstrate that linguistic variation in human--computer interaction is about more than simply recognition accuracy. There is a two-way process of stereotype enforcement: speakers adapt their own speech according to what they believe technologies are capable of recognising, while synthetic voices are themselves evaluated through established social meanings and stereotypes attached to accent. The following sections consider these two aspects before bringing them together as complementary dimensions of sociolinguistic interaction with voice technologies.

\subsection{Accommodation to Perceived Intelligibility}
\label{sec5.1}

When voice technologies failed to recognise participants' ordinary, naturalistic speech, users reported speaking more slowly, enunciating more carefully, repeating or reformulating utterances, avoiding local vocabulary, and shifting towards an accent perceived as more mainstream. These adjustments resemble accommodation practices participants reported using with non-local speakers and in professional or institutional settings. With voice technologies, speakers draw on an existing repertoire of linguistic accommodation in order to increase perceived intelligibility, i.e. what they believe the system is capable of recognising.

This accommodation mostly stemmed from social stereotypes about regional and mainstream accents. Participants described successful machine-directed speech using terms such as ``proper English'', ``neutral'', ``posher'', ``Southern'', or simply ``less Geordie''. Such descriptions reveal participants' model of a successful human--computer interaction, in which regional dialect is considered to be less compatible with voice technology than more mainstream speech.

Voice technology and, in particular, ASR bias may consequently have behavioural effects. Repeated recognition difficulties can encourage speakers to suppress regional features and converge with dialects they perceive as technologically privileged. This places the burden of successful interaction partly on users, who must diagnose the presumed source of recognition failure and modify their speech accordingly. Accommodation in this context can be quite labour-intensive: speakers adjust pronunciation, speech rate, lexical choices, or formulation to compensate for the perceived limitations of the system. When these attempts remain unsuccessful, participants may eventually give up on the interaction altogether.

\subsection{Regional VA Voices: Familiarity vs. Authority}
\label{sec5.2}

The evaluations of the Geordie and SSBE VA voices demonstrate a complementary process, where interaction and perception of a technology is influenced by existing social stereotypes. Participants associated the Geordie voice particularly strongly with familiarity, warmth, personality, and local belonging, whereas the SSBE voice was more often characterised in terms of neutrality and clarity. At the same time, participants frequently anticipated that a Geordie-speaking system could be perceived as less professional, credible, or authoritative. These evaluations resemble well-known sociolinguistic distinctions between the social attractiveness of regional varieties and the status associated with standard speech \citep{stewart1985accent,coupland2007ideologised}. Previous studies have also found mixed evaluations of accented VA voices, with \citet{walker2026false} reporting both familiarity and concerns about trustworthiness and intelligence among Southern US listeners, and \citet{pycha2026perceived} finding that effects on perceived credibility varied across non-native English accents. In sum, VA voices can evoke social meanings associated with human accents, although these associations are not necessarily reproduced in the same way.

Interestingly, participants recognised that these evaluations reflected socially learned biases rather than objective differences in the quality of information provided by the system. This was especially clear in discussions of medical or professional advice: participants acknowledged that the accent of a VA or AI system should have no influence on the reliability of its information, but they still anticipated that a regional voice might be taken less seriously, particularly in such high-stake contexts. 

Regional representation also introduced questions of authenticity. Simply increasing the number or salience of local features did not necessarily make a voice sound more authentically Geordie. Some participants interpreted strongly localised features as stereotypical, exaggerated, or performative, despite the voice having been recorded by a Geordie speaker. This aligns with \citet{jones2026animating}'s argument that exaggerated linguistic features can make synthetic voices sound stereotypical, as well as \citet{michel2025its}'s finding that familiar AI voices can evoke a sense of belonging but may also be perceived as mocking or unrepresentative. Effective regional representation will have to balance recognisable local identity with naturalness and intelligibility while avoiding caricature.

\subsection{Convergence and Divergence in Human--Computer Accommodation}
\label{sec5.3}

The findings demonstrate two ways in which sociolinguistic inequalities can emerge in human--computer interaction: users may suppress dialectal features to improve perceived intelligibility, while VA voices are evaluated through existing accent stereotypes. However, accommodation need not always involve movement towards a standard variety. One participant described the Geordie VA as prompting precisely the opposite response:

\begin{quote}
``Very natural. It was funny hearing the Geordie accent. It made me feel perhaps I should speak broader as well. It would make a Geordie feel at home.'' --- P018 (M, 60+)
\end{quote}

There is a distinction between convergence (or divergence) prompted by recognition difficulties and convergence prompted by social alignment \citep{giles1991language, giles2023communication}. Human--computer accommodation may therefore involve movement either away from or towards dialectal speech, depending on whether speakers are orienting primarily towards the perceived linguistic capabilities of the technology or towards the social identity it represents.

\subsection{Representation or Being Understood?}
\label{sec5.4}

The findings raise a question about what linguistic inclusion in voice technology should prioritise. On the one hand, participants valued aspects of regional representation, with some suggesting that hearing regional voices in technological and public contexts could increase the visibility and acceptance of Newcastle English. On the other hand, participants did not uniformly agree that hearing their own accent from a VA was necessary---in fact, it could even be perceived as mockery. Contrary to that, being understood by a VA was considered to be far more important for a comfortable communication.

The assumption that greater linguistic representation is necessarily desirable also has parallels in language documentation and revitalisation. As \citet{gerdts2010beyond} observes, the priorities of linguists and language communities do not necessarily align: while linguists may prioritise documentation and academic outputs, communities may place greater value on immediate practical needs and maintaining control over how their language is used and represented. A similar principle applies to speech technology. The inclusion of a regional variety should not in itself be taken as evidence that the technology meets the needs or preferences of its speakers. Speakers' own priorities should inform decisions about how their varieties are incorporated into such systems.

For our participants, this suggests prioritising technologies that recognise habitual regional speech without requiring speakers to suppress dialectal features. This is consistent with \citet{blaschke2024dialect}, who found that German dialect speakers generally preferred technologies that process dialectal input over those that produce dialectal output, particularly in the case of VAs. Comparable user-centred arguments have been made for health communication, where linguistic responsiveness is argued to require engagement with the varieties of the communities addressed \citep{Davisetal_fromEnglishtoEnglishes_JMIR_2024}. Regional synthetic voices clearly provide meaningful social benefits in terms of representation, familiarity, and visibility, but it is not useful without, first and foremost, high-quality recognition of regional speech. Ideally, speech technologies should achieve both: users should be able to hear diverse voices represented by these systems while also being confident that their own way of speaking will be understood.

\subsection{Limitations and Future Research}
\label{sec5.5}

The accommodation patterns examined here are based on participants' metalinguistic reports rather than direct measurements of their speech production. The findings demonstrate how participants perceive and describe their accommodation, but cannot establish whether the reported strategies correspond directly to measurable changes in their speech. This distinction is particularly important because some adjustments, such as enunciation and accent standardisation, overlapped considerably in participants' accounts.

Future acoustic analysis should establish whether reported strategies such as standardising, slowing down, or enunciating correspond to measurable phonetic and prosodic changes during interaction with voice technologies \citep[cf.][]{szekely2025will}. A related limitation concerns the VA voices themselves. Perceived attributes such as creditability or authority may depend on acoustic properties of the voice beyond accent \citep[e.g.,][]{Steffens2024CredibleVoice}. Because the Geordie and SSBE stimuli were produced by one speaker, accent may be partly confounded with such properties. Future work should disentangle them, for example by comparing the acoustics of the stimuli. Comparative analysis across dialectal and mainstream voice conditions could further determine whether the system's accent affects the direction or magnitude of accommodation. Such work would make it possible to compare speakers' metalinguistic accounts with their actual linguistic behaviour and to establish more directly whether regional synthetic voices encourage convergence towards dialectal speech.

\section{Conclusion}
\label{sec6}

This study shows that linguistic bias in voice technologies extends beyond differences in recognition accuracy. Newcastle English speakers reported modifying their speech in response to perceived technological limitations, often by reducing dialectal features and moving towards forms they considered more standard or intelligible to the system. At the same time, evaluations of the Geordie and SSBE VA voices reflected existing sociolinguistic attitudes: the regional voice was associated with familiarity, warmth, and local belonging, but also with concerns about authority and professionalism, while the SSBE voice was perceived as more neutral, clear, and credible. These findings demonstrate how voice technologies can reproduce existing sociolinguistic inequalities through both users’ expectations about how they need to speak and the social meanings attached to the voices that technologies produce.

Importantly, linguistic inclusion cannot be reduced to simply increasing the range of accents represented by a system. Designing inclusive voice technologies requires not only reducing disparities in recognition accuracy, but also understanding what speakers themselves want from these technologies and how their preferences are informed by wider sociolinguistic attitudes \citep[see][for a recent overview]{davis2025black}. Human–AI interaction provides a new context in which to examine established sociolinguistic processes: patterns of accommodation, language attitudes, and linguistic stereotyping do not disappear when the interlocutor is artificial. Sociolinguistics is therefore well positioned to bridge the study of linguistic behaviour and the development of language technologies, bringing users’ experiences and linguistic priorities into decisions about what more inclusive voice AI should look and sound like.

\section{Acknowledgements}
We acknowledge the Hochschulinterne Forschungsförderung (HiFF) at Hochschule Düsseldorf. Together with our research assistants Paula Thees and Vincent Reichmann, this support made the work possible as a collaborative team effort. We are also grateful to Prof. Karen Corrigan and Dr. Adam Mearns for their support during data collection.

\section{Author contributions: CRediT}
We follow the CRediT taxonomy\footnote{\url{https://credit.niso.org/}}. Conceptualisation: DS, KT, JS; Data curation: DS; Formal Analysis: DS; Funding acquisition: JS, KT, DS; Investigation: DS; Methodology: DS; Resources: KT, JS; Software: JS, KT, DS; Visualisation: DS, KT, JS; and Writing – original draft: DS, review \& editing: DS, KT, JS.

\section{Data availability}
Due to the qualitative nature of the data and to protect participant anonymity, the data are not publicly available. Anonymised data may be made available upon reasonable request. Researchers interested in accessing the data should contact the first author. The R script and data used to analyse participants' reported
accommodation strategies are available on Zenodo:
\url{https://doi.org/10.5281/zenodo.23685953}.

\section{Declaration of generative AI and AI-assisted technologies in the manuscript preparation process.}
During the preparation of this work the author(s) used ChatGPT for language refinement and readability enhancement. After using this tool/service, the author(s) reviewed and edited the content as needed and take(s) full responsibility for the content of the published article.

\appendix

\section{Participant Distribution}
\label{app:participants}

Participants were allocated to either the Geordie VA condition or the SSBE control condition. As the Geordie VA constituted the primary experimental condition, a larger proportion of participants were assigned to this condition, while the SSBE condition served as a comparison baseline.

\begin{table}[H]
    \centering
    \caption{Participant distribution by age, gender, and VA system they interacted with.}
    \label{tab:participant_distribution}
    \begin{tabular}{llrrr}
        \hline
        \textbf{Age} & \textbf{Gender} & \textbf{Geordie} & \textbf{SSBE} & \textbf{Total} \\
        \hline
        18--29 & Female & 6 & 3 & 9 \\
               & Male   & 7 & 3 & 10 \\
        30--44 & Female & 7 & 2 & 9 \\
               & Male   & 3 & 2 & 5 \\
        45--59 & Female & 1 & 3 & 4 \\
               & Male   & 3 & 3 & 6 \\
        60+    & Female & 4 & 3 & 7 \\
               & Male   & 5 & 3 & 8 \\
        \hline
        \textbf{Total} & & \textbf{36} & \textbf{22} & \textbf{58} \\
        \hline
    \end{tabular}
\end{table}

\section{Post-hoc Semi-Structured Interview}
\label{app:interview}

The following questions formed the post-hoc semi-structured interview conducted with participants after the experimental interaction.

\subsection{Initial Question}

\begin{itemize}
    \item What do you think this experiment was about?
\end{itemize}

\subsection{Perceived System Performance}

\begin{itemize}
    \item How would you rate the accuracy of the VA's responses to your queries?
    \item Did the system understand your accent and dialect without significant issues?
\end{itemize}

\subsection{Interaction Comfort and Naturalness}

\begin{itemize}
    \item Did you feel comfortable speaking with the VA?
    \item How natural did the interaction with the VA feel to you?
\end{itemize}

\subsection{Speech Adaptation Strategies}

\begin{itemize}
    \item Did you find yourself altering your speech in any way during the interaction? If yes, how?
    \item What strategies did you use to make yourself better understood by the VA?
\end{itemize}

\subsection{Perceived System Bias and Inclusivity}

\begin{itemize}
    \item Did you feel that the VA was biased in any way against your dialect or manner of speaking?
    \item How inclusive do you think the VA is for speakers of your dialect?
\end{itemize}

\subsection{Emotional and Behavioral Responses}

\begin{itemize}
    \item How did interacting with the VA make you feel (e.g., frustrated, satisfied)?
    \item Would you prefer interacting with the dialect-sensitive system or a more standard system? Why?
    \item How would you describe the VA (e.g., efficient, unpleasant, polite)?
\end{itemize}

\subsection{Suggestions for Improvement}

\begin{itemize}
    \item What improvements would you suggest for the VA based on your experience?
    \item Are there specific areas where the system failed to meet your expectations?
\end{itemize}

\subsection{Previous Experience with VAs}

\begin{itemize}
    \item What is your general experience with VAs?
    \item How often do you use them?
    \item Have you had to adapt your speech when using technology? How?
    \item What is your general attitude towards technology?
\end{itemize}

\subsection{Demographic and Sociolinguistic Background}

The following demographic and background information was elicited:

\begin{itemize}
    \item Age
    \item Gender
    \item Location
    \item Educational background
    \item Job status
\end{itemize}

\subsection{Regional Orientation and Community}

\begin{itemize}
    \item How connected do you feel to the community where your dialect is spoken?
    \item Do you like Newcastle?
    \item Have you ever thought about moving away?
\end{itemize}

\subsection{Dialect, Identity, and Representation}

\begin{itemize}
    \item What dialect or variety of English do you consider your primary one?
    
    \item How do you feel about the way your dialect is represented or perceived in media and technology?
    
    \item Do you believe your dialect should be more widely supported by technology?
    
    \item How important is your dialect as a part of your identity?
    
    \item Have you ever felt the need to modify your dialect in different social contexts?
    
    \item Are there any unique features in Newcastle English?
    
    \item How would you describe this accent?
    
    \item Can you give me an example of language (words, pronunciation, or phrases) that is authentically Newcastle?
    
    \item Can you think of any examples that the VA failed to produce?
\end{itemize}

\bibliographystyle{apalike} 
\bibliography{Tang}

@PREAMBLE{"\def\authornoop#1{}"}

@article{TianGriffithPriceBoyerTang_2024_LinguisticAlignment_LAS,
author = {Xiaoyi Tian and Amanda E. Griffith and Zane Price and Kristy Elizabeth Boyer and Kevin Tang},
title ={Investigating Linguistic Alignment in Collaborative Dialogue: A Study of Syntactic and Lexical Patterns in Middle School Students},
journal = {Language and Speech},
volume = {0},
number = {0},
pages = {00238309241234565},
year = {2024},
doi = {10.1177/00238309241234565}
}

@inproceedings{friedberg2012lexical,
  author={Friedberg, Heather and Litman, Diane and Paletz, Susannah B. F.},
  booktitle={{2012 IEEE Spoken Language Technology Workshop (SLT)}}, 
  title={Lexical entrainment and success in student engineering groups}, 
  year={2012},
  volume={},
  number={},
  pages={404-409},
  doi={10.1109/SLT.2012.6424258}
  }

@inproceedings{hirschberg2008high,
  title={High frequency word entrainment in spoken dialogue},
  author={Hirschberg, Julia Bell and Nenkova, Ani and Gravano, Agustin},
  year={2008},
    booktitle={{Proceedings of the 46th Annual Meeting of the
Association for Computational Linguistics on Human
Language Technologies}},  
  pages={169--172}
}

@inproceedings{mitchell2012strangers,
  title={From strangers to partners: Examining convergence within a longitudinal study of task-oriented dialogue},
  author={Mitchell, Christopher and Boyer, Kristy and Lester, James},
  booktitle={Proceedings of the 13th Annual Meeting of the Special Interest Group on Discourse and Dialogue},
  pages={94--98},
  year={2012}
}

@phdthesis{davis2025black,
  title={Black Mecha (linguistics): Addressing Antiblackness in Automated Speech Recognition Systems},
  author={Davis, Alexis Ciara},
  year={2025},
  school={University of Florida}
}

@Article{Davisetal_fromEnglishtoEnglishes_JMIR_2024,
  author   = {Davis, Alexis and Martin, Joshua and Cooks, Eric and Vilaro, Melissa and Wilson-Howard, Danyell and Tang, Kevin and Raup Krieger, Janice},
  title    = {From {E}nglish to ``{E}nglishes'': A Process Perspective on Enhancing the Linguistic Responsiveness of Culturally Tailored Cancer Prevention Interventions},
  journal  = {Journal of Participatory Medicine},
  year     = {2024},
  volume   = {16},
  month    = {Dec},
  pages    = {e57528},
  issn     = {2152-7202},
  doi      = {10.2196/57528},
  day      = {19},
}

@InProceedings{MojaradTang_2026_Probing_Interspeech,
  author    = {Hamid Mojarad and Kevin Tang},
  title     = {{Layer-wise Probing of wav2vec 2.0 and {W}hisper for Consonant Cluster Reduction in {A}frican {A}merican {E}nglish}},
  booktitle = {{Interspeech 2026 [Long Track]}},
  year      = {2026},
  pages     = {2251--2260},
  doi       = {10.21437/Interspeech.2026-808},
  issn      = {2958-1796},
}

@Article{serditova2026sociolinguistic,
  author  = {Serditova, Dana and Tang, Kevin},
  title   = {A {S}ociolinguistic {A}nalysis of {A}utomatic {S}peech {R}ecognition {B}ias in {N}ewcastle {E}nglish},
  journal = {Speech Communication},
  year    = {2026},
  note    = {Accepted for publication},
  doi     = {10.48550/arXiv.2603.24549},
  url     = {https://doi.org/10.48550/arXiv.2603.24549},
}

@article{grieve2025sociolinguistic,
 TITLE={The sociolinguistic foundations of language modeling},
  author={Grieve, Jack and Bartl, Sara and Fuoli, Matteo and Grafmiller, Jason and Huang, Weihang and Jawerbaum, Alejandro and Murakami, Akira and Perlman, Marcus and Roemling, Dana and Winter, Bodo},
  journal={Frontiers in Artificial Intelligence},
  volume={7},
  pages={1472411},
  year={2025},
  DOI={10.3389/frai.2024.1472411},
ISSN={2624-8212},
  publisher={Frontiers Media SA}
}

@ARTICLE{CohnZellou2021,
  
AUTHOR={Cohn, Michelle and Zellou, Georgia},   
	 
TITLE={Prosodic Differences in Human- and {A}lexa-Directed Speech, but Similar Local Intelligibility Adjustments},      
	
JOURNAL={Frontiers in Communication},      
	
VOLUME={6},           
	
YEAR={2021},      
	  
URL={https://www.frontiersin.org/articles/10.3389/fcomm.2021.675704},       
	
DOI={10.3389/fcomm.2021.675704},      
	
ISSN={2297-900X}
}

@inproceedings{martin2020understanding,
  title={Understanding Racial {D}isparities in {A}utomatic {S}peech {R}ecognition: {T}he {C}ase of {H}abitual "be"},
  author={Martin, Joshua L and Tang, Kevin},
  booktitle={Interspeech},
  pages={626--630},
  year={2020}
}

@article{Johnson2024,
  author    = {Johnson, Alexander and Shankar, Natarajan Balaji and Ostendorf, Mari and Alwan, Abeer},
  title     = {An exploratory study on dialect density estimation for children and adult's {A}frican {A}merican {E}nglish},
  journal   = {The Journal of the Acoustical Society of America},
  year      = {2024},
  volume    = {155},
  number    = {4},
  pages     = {2836--2848},
  month     = {Apr},
  doi       = {10.1121/10.0025771},
  pmid      = {38682915},
  pmcid     = {PMC11175658},
  publisher = {Acoustical Society of America}
}

@article{koenecke2020racial,
  title={Racial {D}isparities in {A}utomated {S}peech {R}ecognition},
  author={Koenecke, Allison and Nam, Andrew and Lake, Emily and Nudell, Joe and Quartey, Minnie and Mengesha, Zion and Toups, Connor and Rickford, John R and Jurafsky, Dan and Goel, Sharad},
  journal={Proceedings of the National Academy of Sciences},
  volume={117},
  number={14},
  pages={7684--7689},
  year={2020},
  publisher={National Acad Sciences}
}

@inproceedings{heuser24_interspeech,
  title     = {{Quantification of stylistic differences in human- and ASR-produced transcripts of African American English}},
  author    = {Annika Heuser and Tyler Kendall and Miguel {del Rio} and Quinn McNamara and Nishchal Bhandari and Corey Miller and Migüel Jetté},
  year      = {2024},
  booktitle = {{Interspeech 2024}},
  pages     = {4538--4542},
  doi       = {10.21437/Interspeech.2024-2300},
  issn      = {2958-1796},
}

@article{mengesha2021don,
AUTHOR={Mengesha, Zion  and Heldreth, Courtney  and Lahav, Michal  and Sublewski, Juliana  and Tuennerman, Elyse },
           
TITLE={“I don’t Think These Devices are Very Culturally Sensitive.” — Impact of Automated Speech Recognition Errors on {A}frican {A}mericans},
          
JOURNAL={Frontiers in Artificial Intelligence},
          
VOLUME={4},
  pages={725911},
YEAR={2021},
  
URL={https://www.frontiersin.org/journals/artificial-intelligence/articles/10.3389/frai.2021.725911},
  
DOI={10.3389/frai.2021.725911},
  
ISSN={2624-8212}}

@INPROCEEDINGS{liu2022towards,
  author={Liu, Chunxi and Picheny, Michael and Sarı, Leda and Chitkara, Pooja and Xiao, Alex and Zhang, Xiaohui and Chou, Mark and Alvarado, Andres and Hazirbas, Caner and Saraf, Yatharth},
  booktitle={ICASSP 2022 - 2022 IEEE International Conference on Acoustics, Speech and Signal Processing (ICASSP)}, 
  title={Towards Measuring Fairness in Speech Recognition: Casual Conversations Dataset Transcriptions}, 
  year={2022},
  volume={},
  number={},
  pages={6162-6166},
  doi={10.1109/ICASSP43922.2022.9747501}}

@article{feng2024towards,
  title={Towards {I}nclusive {A}utomatic {S}peech {R}ecognition},
  author={Feng, Siyuan and Halpern, Bence Mark and Kudina, Olya and Scharenborg, Odette},
  journal={Computer Speech \& Language},
  volume={84},
  pages={101567},
  year={2024},
  publisher={Elsevier}
}

@ARTICLE{zellou2021age,
    
AUTHOR={Zellou, Georgia  and Cohn, Michelle  and Ferenc Segedin, Bruno },
           
TITLE={Age- and Gender-Related Differences in Speech Alignment Toward Humans and Voice-{AI}},
          
JOURNAL={Frontiers in Communication},
VOLUME={5},
pages={600361},
YEAR={2021},
  
URL={https://www.frontiersin.org/journals/communication/articles/10.3389/fcomm.2020.600361},
  
DOI={10.3389/fcomm.2020.600361},
  
ISSN={2297-900X}}

@article{wassink2022uneven,
  title={Uneven {S}uccess: {A}utomatic {S}peech {R}ecognition and {E}thnicity-{R}elated {D}ialects},
  author={Wassink, Alicia Beckford and Gansen, Cady and Bartholomew, Isabel},
  journal={Speech Communication},
  volume={140},
  pages={50--70},
  year={2022},
  publisher={Elsevier}
}

@article{gessinger2021phonetic,
  title={Phonetic accommodation in interaction with a virtual language learning tutor: A {Wizard-of-Oz} study},
  author={Gessinger, Iona and M{\"o}bius, Bernd and Le Maguer, S{\'e}bastien and Raveh, Eran and Steiner, Ingmar},
  journal={Journal of Phonetics},
  volume={86},
  pages={101029},
  year={2021},
  publisher={Elsevier}
}

@article{strycharczuk2020general,
AUTHOR={Strycharczuk, Patrycja  and López-Ibáñez, Manuel  and Brown, Georgina  and Leemann, Adrian },
           
TITLE={General {N}orthern {E}nglish. Exploring Regional Variation in the North of {E}ngland With Machine Learning},
          
JOURNAL={Frontiers in Artificial Intelligence},
          
VOLUME={3},
  number={48},
YEAR={2020},
  
URL={https://www.frontiersin.org/journals/artificial-intelligence/articles/10.3389/frai.2020.00048},
  
DOI={10.3389/frai.2020.00048},
  
ISSN={2624-8212}}

@book{giles1991language,
  title={Language: Contexts and consequences.},
  author={Giles, Howard and Coupland, Nikolas},
  year={1991},
  publisher={Thomson Brooks/Cole Publishing Co},  
  address   = {Pacific Grove, CA},
}

@article{Steffens2024CredibleVoice,
  title={How does a credible voice sound?},
  author={Steffens, Jochen and Bl{\"a}ttermann, Patrick and Sattler, Maximilian and Tang, Kevin},
  journal={Journal of the Acoustical Society of America Express Letters},
  year={forthcoming},
  note={Forthcoming publication}
}

@article{cohn2024african,
    author = {Cohn, Michelle and Mengesha, Zion and Lahav, Michal and Heldreth, Courtney},
    title = {African {A}merican {E}nglish speakers’ pitch variation and rate adjustments for imagined technological and human addressees},
    journal = {JASA Express Letters},
    volume = {4},
    number = {4},
    pages = {047601},
    year = {2024},
    month = {04},
    issn = {2691-1191},
    doi = {10.1121/10.0025484},
    url = {https://doi.org/10.1121/10.0025484},
    eprint = {https://pubs.aip.org/asa/jel/article-pdf/doi/10.1121/10.0025484/19910179/047601_1_10.0025484.pdf},
}

@inproceedings{ngueajio2022hey,
  title={Hey {ASR} {S}ystem! {W}hy {A}ren’t {Y}ou {M}ore {I}nclusive? {A}utomatic {S}peech {R}ecognition {S}ystems’ {B}ias and {P}roposed {B}ias {M}itigation {T}echniques. {A} {L}iterature {R}eview},
  author={Ngueajio, Mikel K and Washington, Gloria},
  booktitle={International Conference on Human-Computer Interaction},
  pages={421--440},
  year={2022},
  organization={Springer}
}

@inproceedings{markl2022language,
  title={Language {V}ariation and {A}lgorithmic {B}ias: {U}nderstanding algorithmic bias in {B}ritish {E}nglish {A}utomatic {S}peech {R}ecognition},
  author={Markl, Nina},
  booktitle={Proceedings of the 2022 ACM Conference on Fairness, Accountability, and Transparency},
  pages={521--534},
  year={2022}
}

@inproceedings{sanabria2023edinburgh,
  title={The {E}dinburgh {I}nternational {A}ccents of {E}nglish {C}orpus: Towards the democratization of {E}nglish {ASR}},
  author={Sanabria, Ramon and Bogoychev, Nikolay and Markl, Nina and Carmantini, Andrea and Klejch, Ondrej and Bell, Peter},
  booktitle={ICASSP 2023-2023 IEEE International Conference on Acoustics, Speech and Signal Processing (ICASSP)},
  pages={1--5},
  year={2023},
  organization={IEEE}
}

@article{zellou2024linguistic,
  title={Linguistic analysis of human-computer interaction},
  author={Zellou, Georgia and Holliday, Nicole},
  journal={Frontiers in Computer Science},
  volume={6},
  pages={1384252},
  year={2024},
  publisher={Frontiers Media SA}
}

@article{lai2025dialect,
  title={Dialect {B}ias in {A}utomatic {S}peech {R}ecognition: Analysis of {A}ppalachian {E}nglish},
  author={Lai, Li-Fang and van Hell, Janet G and Lipski, John},
  journal={American Speech: A Quarterly of Linguistic Usage},
  volume={100},
  number={2},
  pages={190--207},
  year={2025},
  publisher={Duke University Press}
}

@article{pearce2012folk,
  title={Folk {A}ccounts of {D}ialect {D}ifferences in {T}yne and {W}ear},
  author={Pearce, Michael},
  journal={Dialectologia et Geolinguistica},
  volume={20},
  number={1},
  pages={5--25},
  year={2012},
  publisher={De Gruyter}
}

@phdthesis{burbano2008language,
  title={Language variation and identity in {S}underland},
  author={Burbano-Elizondo, Lourdes},
  year={2008},
  school={University of Sheffield}
}

@article{montgomery2025feature,
  title={Feature attention and accent recognition: human listeners’ responses to five {N}orthern {E}nglish accents},
  author={Montgomery, Chris and Vriesendorp, Hielke and Walker, Gareth},
  journal={Frontiers in Psychology},
  volume={16},
  pages={1613018},
  year={2025},
  publisher={Frontiers Media SA}
}

@article{beal2009enregisterment,
  title={Enregisterment, commodification, and historical context:“{G}eordie” versus “{S}heffieldish”},
  author={Beal, Joan C},
  journal={American speech},
  volume={84},
  number={2},
  pages={138--156},
  year={2009},
  publisher={Duke University Press}
}

@inbook{pearce2017linguistic,
url = {https://doi.org/10.1515/9783110450903-004},
title = {4. The linguistic landscape of north-east {E}ngland},
booktitle = {Perspectives on Northern Englishes},
author = {Michael Pearce},
editor = {Sylvie Hancil and Joan C. Beal},
publisher = {De Gruyter Mouton},
address = {Berlin, Boston},
pages = {61--82},
doi = {doi:10.1515/9783110450903-004},
isbn = {9783110450903},
year = {2017},
lastchecked = {2026-10-25}
}

@article{pearce2009perceptual,
  title={A {P}erceptual {D}ialect {M}ap of {N}orth {E}ast {E}ngland},
  author={Pearce, Michael},
  journal={Journal of English Linguistics},
  volume={37},
  number={2},
  pages={162--192},
  year={2009},
  publisher={Sage Publications Sage CA: Los Angeles, CA}
}

@inproceedings{torgbi2025adapting,
    title = "Adapting {W}hisper for Regional Dialects: Enhancing Public Services for Vulnerable Populations in the {U}nited {K}ingdom",
    author = "Torgbi, Melissa  and
      Clayman, Andrew  and
      Speight, Jordan J.  and
      Tayyar Madabushi, Harish",
    editor = "Scherrer, Yves  and
      Jauhiainen, Tommi  and
      Ljube{\v{s}}i{\'c}, Nikola  and
      Nakov, Preslav  and
      Tiedemann, Jorg  and
      Zampieri, Marcos",
    booktitle = "Proceedings of the 12th Workshop on {NLP} for Similar Languages, Varieties and Dialects",
    month = jan,
    year = "2025",
    address = "Abu Dhabi, UAE",
    publisher = "Association for Computational Linguistics",
    url = "https://aclanthology.org/2025.vardial-1.4/",
    pages = "29--38"
}

@article{labov1963social,
  title={The {S}ocial {M}otivation of a {S}ound {C}hange},
  author={Labov, William},
  journal={Word},
  volume={19},
  number={3},
  pages={273--309},
  year={1963},
  publisher={Taylor \& Francis}
}

@book{labov1972sociolinguistic,
  author    = {Labov, William},
  title     = {Sociolinguistic Patterns},
  year      = {1972},
  address   = {Philadelphia},
  publisher = {University of Pennsylvania Press},
  series    = {Conduct and Communication},
  number    = {4},
  isbn      = {9780812210408}
}

@article{bell1984language,
  title={Language {S}tyle as {A}udience {D}esign},
  author={Bell, Allan},
  journal={Language in society},
  volume={13},
  number={2},
  pages={145--204},
  year={1984},
  publisher={Cambridge University Press}
}

@INPROCEEDINGS{jahan2025unveiling,
  author={Jahan, Maliha and Mazumdar, Priyam and Thebaud, Thomas and Hasegawa-Johnson, Mark and Villalba, Jesús and Dehak, Najim and Moro-Velazquez, Laureano},
  booktitle={ICASSP 2025 - 2025 IEEE International Conference on Acoustics, Speech and Signal Processing (ICASSP)}, 
  title={Unveiling Performance Bias in {ASR} Systems: {A} Study on Gender, Age, Accent, and More}, 
  year={2025},
  volume={},
  number={},
  pages={1-5},
  doi={10.1109/ICASSP49660.2025.10887767}}

@inproceedings{lai2023exploring,
  title     = {{Exploring Sources of Racial Bias in Automatic Speech Recognition through the Lens of Rhythmic Variation}},
  author    = {Li-Fang Lai and Nicole Holliday},
  year      = {2023},
  booktitle = {{Interspeech 2023}},
  pages     = {1284--1288},
  doi       = {10.21437/Interspeech.2023-159},
  issn      = {2958-1796},
}

@article{halpern2022low,
  title={Low-{R}esource {A}utomatic {S}peech {R}ecognition and {E}rror {A}nalyses of {O}ral {C}ancer {S}peech},
  author={Halpern, Bence Mark and Feng, Siyuan and van Son, Rob and van den Brekel, Michiel and Scharenborg, Odette},
  journal={Speech Communication},
  volume={141},
  pages={14--27},
  year={2022},
  publisher={Elsevier}
}

@inproceedings{serditova25_interspeech,
  title     = {{Automatic Speech Recognition Biases in Newcastle English: an Error Analysis}},
  author    = {Dana Serditova and Kevin Tang and Jochen Steffens},
  year      = {2025},
  booktitle = {{Interspeech 2025}},
  pages     = {3204--3208},
  doi       = {10.21437/Interspeech.2025-1973},
  issn      = {2958-1796},
}

@article{kelly2022sociolinguistics,
  title={Sociolinguistics in an {I}ncreasingly {T}echnologized {R}eality},
  author={Kelly-Holmes, Helen},
  journal={Sociolinguistica},
  volume={36},
  number={1-2},
  pages={99--110},
  year={2022},
  publisher={De Gruyter}
}

@inproceedings{szekely2025will,
    title = "Will {AI} shape the way we speak? The emerging sociolinguistic influence of synthetic voices",
    author = "Szekely, Eva  and
      Miniota, Jura  and
      Hejn{\'a}, M{\'i}{\v{s}}a (Michaela)",
    editor = "Torres, Maria Ines  and
      Matsuda, Yuki  and
      Callejas, Zoraida  and
      del Pozo, Arantza  and
      D{'}Haro, Luis Fernando",
    booktitle = "Proceedings of the 15th International Workshop on Spoken Dialogue Systems Technology",
    month = may,
    year = "2025",
    address = "Bilbao, Spain",
    publisher = "Association for Computational Linguistics",
    url = "https://aclanthology.org/2025.iwsds-1.37/",
    pages = "335--340",
    ISBN = "979-8-89176-248-0"
}

@inproceedings{butler2019exploration,
author = {Butler, Janine and Trager, Brian and Behm, Byron},
title = {Exploration of Automatic Speech Recognition for Deaf and Hard of Hearing Students in Higher Education Classes},
year = {2019},
isbn = {9781450366762},
publisher = {Association for Computing Machinery},
address = {New York, NY, USA},
url = {https://doi.org/10.1145/3308561.3353772},
doi = {10.1145/3308561.3353772},
booktitle = {Proceedings of the 21st International ACM SIGACCESS Conference on Computers and Accessibility},
pages = {32–42},
numpages = {11},
location = {Pittsburgh, PA, USA},
series = {ASSETS '19}
}

@inproceedings{cumbal2024you,
  title     = {“You don’t understand me!”: {C}omparing {ASR} Results for {L}1 and {L}2 Speakers of {S}wedish},
  author    = {Ronald Cumbal and Birger Moell and José Lopes and Olov Engwall},
  year      = {2021},
  booktitle = {{Interspeech 2021}},
  pages     = {4463--4467},
  doi       = {10.21437/Interspeech.2021-2140},
  issn      = {2958-1796},
}

@misc{adedeji2024sound,
      title={The {S}ound of {H}ealthcare: {I}mproving medical transcription {ASR} accuracy with large language models},
      author={Adedeji, Ayo and Joshi, Sarita and Doohan, Brendan},
      year={2024},
      eprint={2402.07658},
      archivePrefix={arXiv},
      primaryClass={cs.CL},
      url={https://arxiv.org/abs/2402.07658}, 
}

@ARTICLE{latif2020speech,
  author={Latif, Siddique and Qadir, Junaid and Qayyum, Adnan and Usama, Muhammad and Younis, Shahzad},
  journal={IEEE Reviews in Biomedical Engineering}, 
  title={Speech Technology for Healthcare: Opportunities, Challenges, and State of the Art}, 
  year={2021},
  volume={14},
  number={},
  pages={342-356},
  doi={10.1109/RBME.2020.3006860}}

@inproceedings{MartinCorpusData2021,
author = {Martin, Joshua L},
title = {Spoken Corpora Data, Automatic Speech Recognition, and Bias Against {A}frican {A}merican Language: The case of Habitual 'Be'},
year = {2021},
isbn = {9781450383097},
publisher = {Association for Computing Machinery},
address = {New York, NY, USA},
url = {https://doi.org/10.1145/3442188.3445893},
doi = {10.1145/3442188.3445893},
booktitle = {Proceedings of the 2021 ACM Conference on Fairness, Accountability, and Transparency},
pages = {284},
numpages = {1},
location = {Virtual Event, Canada},
series = {FAccT '21}
}

@inproceedings{dipto-etal-2025-asr,
    title = "Are {ASR} foundation models generalized enough to capture features of regional dialects for low-resource languages?",
    author = "Dipto, Tawsif Tashwar  and
      Azmol Hossain  and
      Faruque, Rubayet Sabbir  and
      Hassan, Md. Rezuwan  and
      Fatema, Kanij  and
      Shome, Tanmoy  and
      Naswan, Ruwad  and
      Zihad, Md.Foriduzzaman  and
      Anam, Mohaymen Ul  and
      Tasnim, Nazia  and
      Mahmud, Hasan  and
      Hasan, Md Kamrul  and
      Shawon, Md. Mehedi Hasan  and
      Sadeque, Farig  and
      Reasat, Tahsin",
    editor = "Inui, Kentaro  and
      Sakti, Sakriani  and
      Wang, Haofen  and
      Wong, Derek F.  and
      Bhattacharyya, Pushpak  and
      Banerjee, Biplab  and
      Ekbal, Asif  and
      Chakraborty, Tanmoy  and
      Singh, Dhirendra Pratap",
    booktitle = "Proceedings of the 14th International Joint Conference on Natural Language Processing and the 4th Conference of the Asia-Pacific Chapter of the Association for Computational Linguistics",
    month = dec,
    year = "2025",
    address = "Mumbai, India",
    publisher = "The Asian Federation of Natural Language Processing and The Association for Computational Linguistics",
    url = "https://aclanthology.org/2025.ijcnlp-short.17/",
    doi = "10.18653/v1/2025.ijcnlp-short.17",
    pages = "178--188",
    ISBN = "979-8-89176-299-2"
}

@InProceedings{BeraAgarwal2025,
author="Bera, Anup
and Agarwal, Aanchal",
editor="Wang, Chua-Chin
and Sangalang, Ralph Gerard B.",
title="Bias Detection and Mitigation Framework for {ASR} System",
booktitle="7th International Conference on Signal Processing and Information Communications",
year="2025",
publisher="Springer Nature Switzerland",
address="Cham",
pages="13--27",
isbn="978-3-031-89994-2"
}

@incollection{vacher2010complete,
author = {Michel Vacher and Anthony Fleury and Francois Portet and Jean-Francois Serignat and Norbert Noury},
title = {Complete Sound and Speech Recognition System for Health Smart Homes: Application to the Recognition of Activities of Daily Living},
booktitle = {New Developments in Biomedical Engineering},
publisher = {IntechOpen},
address = {London},
year = {2010},
editor = {Domenico Campolo},
chapter = {33},
doi = {10.5772/7596},
url = {https://doi.org/10.5772/7596}
}

@article{hamalainen2015multilingual,
  title={Multilingual {S}peech {R}ecognition for the {E}lderly: {T}he {AALF}red personal life assistant},
  author={H{\"a}m{\"a}l{\"a}inen, Annika and Teixeira, Ant{\'o}nio and Almeida, Nuno and Meinedo, Hugo and Fegy{\'o}, Tibor and Dias, Miguel Sales},
  journal={Procedia Computer Science},
  volume={67},
  pages={283--292},
  year={2015},
  publisher={Elsevier}
}

@inproceedings{markl2021context,
    title = "Context-sensitive evaluation of automatic speech recognition: considering user experience {\&} language variation",
    author = "Markl, Nina  and
      Lai, Catherine",
    editor = "Blodgett, Su Lin  and
      Madaio, Michael  and
      O'Connor, Brendan  and
      Wallach, Hanna  and
      Yang, Qian",
    booktitle = "Proceedings of the First Workshop on Bridging Human{--}Computer Interaction and Natural Language Processing",
    month = apr,
    year = "2021",
    address = "Online",
    publisher = "Association for Computational Linguistics",
    url = "https://aclanthology.org/2021.hcinlp-1.6/",
    pages = "34--40"
}

@inproceedings{green2021automatic,
  title     = {{Automatic Speech Recognition of Disordered Speech: Personalized Models Outperforming Human Listeners on Short Phrases}},
  author    = {Jordan R. Green and Robert L. MacDonald and Pan-Pan Jiang and Julie Cattiau and Rus Heywood and Richard Cave and Katie Seaver and Marilyn A. Ladewig and Jimmy Tobin and Michael P. Brenner and Philip C. Nelson and Katrin Tomanek},
  year      = {2021},
  booktitle = {{Interspeech 2021}},
  pages     = {4778--4782},
  doi       = {10.21437/Interspeech.2021-1384},
  issn      = {2958-1796},
}

@Article{lme2015,
    title = {Fitting Linear Mixed-Effects Models Using {lme4}},
    author = {Douglas Bates and Martin M{\"a}chler and Ben Bolker and
      Steve Walker},
    journal = {Journal of Statistical Software},
    year = {2015},
    volume = {67},
    number = {1},
    pages = {1--48},
    doi = {10.18637/jss.v067.i01},
  }

@Manual{r,
     title = {R: A Language and Environment for Statistical Computing},
     author = {{R Core Team}},
     organization = {R Foundation for Statistical Computing},
     address = {Vienna, Austria},
     year = {2025},
     url = {https://www.R-project.org/},
   }

@article{dichristofano2022global,
  title={Global {P}erformance {D}isparities {B}etween {E}nglish-{L}anguage {A}ccents in {A}utomatic {S}peech {R}ecognition},
  author={DiChristofano, Alex and Shuster, Henry and Chandra, Shefali and Patwari, Neal},
  journal={arXiv preprint arXiv:2208.01157},
  year={2022}
}

@inproceedings{bassignana2025ai,
    title = "The {AI} Gap: How Socioeconomic Status Affects Language Technology Interactions",
    author = "Bassignana, Elisa  and
      Curry, Amanda Cercas  and
      Hovy, Dirk",
    editor = "Che, Wanxiang  and
      Nabende, Joyce  and
      Shutova, Ekaterina  and
      Pilehvar, Mohammad Taher",
    booktitle = "Proceedings of the 63rd Annual Meeting of the Association for Computational Linguistics (Volume 1: Long Papers)",
    month = jul,
    year = "2025",
    address = "Vienna, Austria",
    publisher = "Association for Computational Linguistics",
    url = "https://aclanthology.org/2025.acl-long.914/",
    doi = "10.18653/v1/2025.acl-long.914",
    pages = "18647--18664",
    ISBN = "979-8-89176-251-0"
}

@inbook{guy1988language, place={Cambridge}, title={Language and social class}, booktitle={Linguistics: The Cambridge Survey}, publisher={Cambridge University Press}, author={Guy, Gregory R.}, editor={Newmeyer, Frederick J.Editor}, year={1988}, pages={37–63}}

@inbook{labov1986social, place={Cambridge}, title={The social stratification of (r) in {N}ew {Y}ork {C}ity department stores}, booktitle={The Social Stratification of English in New York City}, publisher={Cambridge University Press}, author={Labov, William}, year={2006}, pages={40–57}}

@article{milroy1993mechanisms,
  title={Mechanisms of {C}hange in {U}rban {D}ialects: {T}he {R}ole of {C}lass, {S}ocial {N}etwork and {G}ender},
  author={Milroy, James and Milroy, Lesley},
  journal={International Journal of Applied Linguistics},
  volume={3},
  number={1},
  pages={57--77},
  year={1993},
  publisher={Wiley Online Library}
}

@inbook{rickford1996regional, place={Cambridge}, series={Cambridge Applied Linguistics}, title={Regional and social variation}, booktitle={Sociolinguistics and Language Teaching}, publisher={Cambridge University Press}, author={Rickford, John R.}, editor={McKay, Sandra Lee and Hornberger, Nancy H.Editors}, year={1995}, pages={151–194}, collection={Cambridge Applied Linguistics}}

@Manual{emmeans,
  title = {emmeans: Estimated Marginal Means, aka Least-Squares Means},
  author = {Russell V. Lenth and Julia Piaskowski},
  year = {2025},
  note = {R package version 2.0.1},
  url = {https://rvlenth.github.io/emmeans/},
}

@inproceedings{harris2024modeling,
    title = "Modeling {G}ender and {D}ialect {B}ias in {A}utomatic {S}peech {R}ecognition",
    author = "Harris, Camille  and
      Mgbahurike, Chijioke  and
      Kumar, Neha  and
      Yang, Diyi",
    editor = "Al-Onaizan, Yaser  and
      Bansal, Mohit  and
      Chen, Yun-Nung",
    booktitle = "Findings of the Association for Computational Linguistics: EMNLP 2024",
    month = nov,
    year = "2024",
    address = "Miami, Florida, USA",
    publisher = "Association for Computational Linguistics",
    url = "https://aclanthology.org/2024.findings-emnlp.890/",
    doi = "10.18653/v1/2024.findings-emnlp.890",
    pages = "15166--15184"
}

@inproceedings{garnerin2021investigating,
    title = "Investigating the {I}mpact of {G}ender {R}epresentation in {ASR} {T}raining {D}ata: a {C}ase {S}tudy on {L}ibrispeech",
    author = "Garnerin, Mahault  and
      Rossato, Solange  and
      Besacier, Laurent",
    editor = "Costa-juss{\`a}, Marta R.  and
      Gonen, Hila  and
      Hardmeier, Christian  and
      Webster, Kellie",
    booktitle = "Proceedings of the 3rd Workshop on Gender Bias in Natural Language Processing",
    month = aug,
    year = "2021",
    address = "Online",
    publisher = "Association for Computational Linguistics",
    url = "https://aclanthology.org/2021.gebnlp-1.10/",
    doi = "10.18653/v1/2021.gebnlp-1.10",
    pages = "86--92"
}

@inproceedings{
zhao2024wildchat,
title={{WildChat}: 1M {ChatGPT} Interaction Logs in the Wild},
author={Wenting Zhao and Xiang Ren and Jack Hessel and Claire Cardie and Yejin Choi and Yuntian Deng},
booktitle={The Twelfth International Conference on Learning Representations},
year={2024},
url={https://openreview.net/forum?id=Bl8u7ZRlbM}
}

@inbook{liang2026wasn,
author = {Liang, Siyu and Wassink, Alicia Beckford},
title = {“This Wasn't Made for Me”: {R}ecentering User Experience and Emotional Impact in the Evaluation of {ASR} Bias},
year = {2026},
isbn = {9798400725968},
publisher = {Association for Computing Machinery},
address = {New York, NY, USA},
url = {https://doi.org/10.1145/3805689.3806448},
booktitle = {Proceedings of the 2026 ACM Conference on Fairness, Accountability, and Transparency},
pages = {7308–7326},
numpages = {19}
}

@inproceedings{nass1994computers,
author = {Nass, Clifford and Steuer, Jonathan and Tauber, Ellen R.},
title = {Computers are social actors},
year = {1994},
isbn = {0897916506},
publisher = {Association for Computing Machinery},
address = {New York, NY, USA},
url = {https://doi.org/10.1145/191666.191703},
doi = {10.1145/191666.191703},
booktitle = {Proceedings of the SIGCHI Conference on Human Factors in Computing Systems},
pages = {72–78},
numpages = {7},
location = {Boston, Massachusetts, USA},
series = {CHI '94}
}

@article{gambino2020building,
  title={Building a {S}tronger {CASA}: {E}xtending the {C}omputers are {S}ocial {A}ctors {P}aradigm},
  author={Gambino, Andrew and Fox, Jesse and Ratan, Rabindra A},
  journal={Human-Machine Communication},
  volume={1},
  pages={71--85},
  year={2020},
  publisher={Communication and Social Robotics Labs Kalamazoo, Michigan}
}

@article{shaw2023revealing,
  title={Revealing perceptual structure through input variation: cross-accent categorization of vowels in five accents of {E}nglish},
  author={Shaw, Jason A and Foulkes, Paul and Hay, Jennifer and Evans, Bronwen G and Docherty, Gerard and Mulak, Karen E and Best, Catherine T},
  journal={Laboratory Phonology},
  volume={14},
  number={1},
  year={2023},
  doi={10.16995/labphon.6436},
  publisher={Open Library of Humanities}
}

@inproceedings{best2015newcastle,
  title={From {N}ewcastle {MOUTH} to {A}ussie ears: {A}ustralians' perceptual assimilation and adaptation for {N}ewcastle {UK} vowels},
  author    = {Catherine T. Best and Jason A. Shaw and Gerard Docherty and Bronwen G. Evans and Paul Foulkes and Jennifer Hay and Jalal Al-Tamimi and Katharine Mair and Karen E. Mulak and Sophie Wood},
  year      = {2015},
  booktitle = {{Interspeech 2015}},
  pages     = {1932--1936},
  doi       = {10.21437/Interspeech.2015-426},
  issn      = {2958-1796},
}

@article{braber2015language,
  title={Language perception in the {E}ast {M}idlands in {E}ngland: {I}nvestigating {E}ast {M}idlands adolescents' perception of language variation in the {UK}},
  author={Braber, Natalie},
  journal={English Today},
  volume={31},
  number={1},
  pages={16--26},
  year={2015},
  publisher={Cambridge University Press}
}

@article{paver2025stereotyped,
TITLE={Stereotyped accent judgements in forensic contexts: listener perceptions of social traits and types of behaviour},
  author={Paver, Alice and Wright, David and Braber, Natalie and Pautz, Nikolas},
  journal={Frontiers in Communication},
  volume={9},
  pages={1462013},
  year={2025},
  publisher={Frontiers Media SA},
  URL={https://www.frontiersin.org/journals/communication/articles/10.3389/fcomm.2024.1462013},
DOI={10.3389/fcomm.2024.1462013},
ISSN={2297-900X},
}

@article{coupland2007ideologised,
  title={Ideologised {V}alues for {B}ritish {A}ccents},
  author={Coupland, Nikolas and Bishop, Hywel},
  journal={Journal of Sociolinguistics},
  volume={11},
  number={1},
  pages={74--93},
  year={2007},
  publisher={Wiley Online Library}
}

@article{ostrand2021its,
  title={It’s {A}lignment {A}ll the {W}ay {D}own, but {N}ot {A}ll the {W}ay {U}p: {S}peakers align on some features but not others within a dialogue},
  author={Ostrand, Rachel and Chodroff, Eleanor},
  journal={Journal of phonetics},
  volume={88},
  pages={101074},
  year={2021},
  publisher={Elsevier}
}

@article{giles2023communication,
  title={Communication {A}ccommodation {T}heory: {P}ast accomplishments, current trends, and future prospects},
  author={Giles, Howard and Edwards, America L and Walther, Joseph B},
  journal={Language Sciences},
  volume={99},
  pages={101571},
  year={2023},
  publisher={Elsevier}
}

@book{labov1966social,
  author    = {William Labov},
  title     = {The Social Stratification of English in New York City},
  year      = {1966},
  publisher = {Center for Applied Linguistics},
  address   = {Washington, DC}
}

@Inbook{trudgill1974social,
author="Trudgill, Peter",
editor="Coupland, Nikolas
and Jaworski, Adam",
title="The Social Differentiation of {E}nglish in {N}orwich",
bookTitle="Sociolinguistics: A Reader",
year="1997",
publisher="Macmillan Education UK",
address="London",
pages="179--184",
isbn="978-1-349-25582-5",
doi="10.1007/978-1-349-25582-5_15",
url="https://doi.org/10.1007/978-1-349-25582-5_15"
}

@book{eckert2000linguistic,
  author    = {Eckert, Penelope},
  title     = {Language Variation as Social Practice:
               The Linguistic Construction of Identity in Belten High},
  year      = {1999},
  publisher = {Blackwell Publishers},
  address   = {Oxford},
  series    = {Language in Society},
  number    = {27},
  isbn      = {9780631186038}
}

@article{cohn2022acoustic,
title = {Acoustic-phonetic properties of {S}iri- and human-directed speech},
journal = {Journal of Phonetics},
volume = {90},
pages = {101123},
year = {2022},
issn = {0095-4470},
doi = {10.1016/j.wocn.2021.101123},
url = {https://www.sciencedirect.com/science/article/pii/S009544702100098X},
author = {Michelle Cohn and Bruno Ferenc Segedin and Georgia Zellou}
}

@book{carmichael2025language, place={Cambridge}, series={Elements in Sociolinguistics}, title={Language and Place}, publisher={Cambridge University Press}, author={Carmichael, Katie and Reed, Paul E.}, year={2025}, collection={Elements in Sociolinguistics}}

@article{monka2020place,
  title={Place attachment and linguistic variation: A quantitative analysis of language and local attachment in a rural village and an urban social housing area},
  author={Monka, Malene and Quist, Pia and Skovse, Astrid Ravn},
  journal={Language in Society},
  volume={49},
  number={2},
  pages={173--205},
  year={2020},
  publisher={Cambridge University Press}
}

@article{stewart1985accent,
  title={Accent and social class effects on status and solidarity evaluations},
  author={Stewart, Mark A and Ryan, Ellen Bouchard and Giles, Howard},
  journal={Personality and social psychology bulletin},
  volume={11},
  number={1},
  pages={98--105},
  year={1985},
  publisher={Sage Publications Sage CA: Thousand Oaks, CA}
}

@inproceedings{blaschke2024dialect,
    title = "What Do Dialect Speakers Want? A Survey of Attitudes Towards Language Technology for {G}erman Dialects",
    author = "Blaschke, Verena  and
      Purschke, Christoph  and
      Schuetze, Hinrich  and
      Plank, Barbara",
    editor = "Ku, Lun-Wei  and
      Martins, Andre  and
      Srikumar, Vivek",
    booktitle = "Proceedings of the 62nd Annual Meeting of the {A}ssociation for {C}omputational {L}inguistics (Volume 2: Short Papers)",
    month = aug,
    year = "2024",
    address = "Bangkok, Thailand",
    publisher = "Association for Computational Linguistics",
    url = "https://aclanthology.org/2024.acl-short.74/",
    doi = "10.18653/v1/2024.acl-short.74",
    pages = "823--841"
}

@book{serditova2026sounds,
  author    = {Serditova, Dana},
  title     = {The Sounds of New Orleans: A Sociophonetic Study},
  year      = {2026},
  publisher = {John Benjamins Publishing Company},
  doi       = {10.1075/veaw.g74},
  isbn      = {9789027237798},
  url       = {https://doi.org/10.1075/veaw.g74}
}

@article{erdocia2026language,
  title={Language in the age of {AI} technology: From human to non-human authenticity, from public governance to privatised assemblages},
  author={Erdocia, Iker and Schneider, Britta and Migge, Bettina},
  journal={Language in Society},
  volume={55},
  number={3},
  pages={523--543},
  year={2026},
  publisher={Cambridge University Press}
}

@incollection{gerdts2010beyond,
  title={Beyond expertise: The role of the linguist in language revitalization programs},
  author={Gerdts, Donna B},
  booktitle={Language documentation: Practice and values},
  pages={173--192},
  year={2010},
  publisher={John Benjamins Publishing Company}
}

@article{walker2026false,
url = {https://doi.org/10.1515/phon-2025-0080},
title = {False memories: subjective and objective responses to an “accented” voice assistant},
author = {Abby Walker and Eliza Quesenberry and Katelyn Morris and Mahmut Sami Gürdal and Shuqi Yu and Koeun Choi and Benjamin Katz and Myounghoon Jeon},
journal = {Phonetica},
doi = {10.1515/phon-2025-0080},
year = {2026},
lastchecked = {2026-10-25}
}

@article{jones2026animating, title={Animating {AI}: Language variation and “enoughness” in voice user interfaces}, volume={46}, DOI={10.1017/S0267190526100312}, journal={Annual Review of Applied Linguistics}, author={Jones, Rodney H.}, year={2026}, pages={34–55}}

@article{pycha2026perceived,
  title={Perceived credibility of native and non-native accents during voice-AI interactions},
  author={Pycha, Anne and Zellou, Georgia},
  journal={Frontiers in Communication},
  volume={11},
  pages={1898373},
  year={2026},
  publisher={Frontiers Media SA}
}

@inproceedings{michel2025its,
  title={“It’s not a representation of me”: examining accent bias and digital exclusion in synthetic {AI} voice services},
  author={Michel, Shira and Kaur, Sufi and Gillespie, Sarah Elizabeth and Gleason, Jeffrey and Wilson, Christo and Ghosh, Avijit},
  booktitle={Proceedings of the 2025 ACM Conference on Fairness, Accountability, and Transparency},
  pages={228--245},
  year={2025}
}

@article{chen2026accommodates,
AUTHOR = {Chen, Pengbo and Guan, Huining and Jeong, Eui Jun},
TITLE = {Who Accommodates Whom? {B}idirectional Linguistic Accommodation and Progressive Interpersonal Convergence in Human–{AI} Conversations},
JOURNAL = {Behavioral Sciences},
VOLUME = {16},
YEAR = {2026},
NUMBER = {5},
ARTICLE-NUMBER = {720},
URL = {https://www.mdpi.com/2076-328X/16/5/720},
PubMedID = {42193597},
ISSN = {2076-328X},
DOI = {10.3390/bs16050720}
}

@inproceedings{scott2026sociophonetic,
    title = "A Sociophonetic Analysis of Racial Bias in Commercial {ASR} Systems Using the {P}acific {N}orthwest {E}nglish Corpus",
    author = "Scott, Michael  and
      Liang, Siyu  and
      Wassink, Alicia  and
      Levow, Gina-Anne",
    editor = "Piperidis, Stelios  and
      Bel, N{\'u}ria  and
      van den Heuvel, Henk  and
      Ide, Nancy  and
      Krek, Simon  and
      Toral, Antonio",
    booktitle = "Proceedings of the Fifteenth Language Resources and Evaluation Conference",
    month = may,
    year = "2026",
    address = "Palma de Mallorca, Spain",
    publisher = "ELRA Language Resource Association",
    url = "https://aclanthology.org/2026.lrec-1.376/",
    doi = "10.63317/4qkgjt7n9ehb",
    pages = "4787--4796"
}




\end{document}